\documentclass{ws-rv975x65}
\usepackage{ws-rv-van}             
\makeindex

\begin{document}

\chapter{Skyrmion Approach to finite density and temperature\label{skyrmion_matter}}

\author[B.-Y. Park and V. Vento]{Byung-Yoon Park$^1$ and Vicente Vento$^2$ }
%\index[aindx]{Author, F.} % or \aindx{Park, B.-Y.}
%\index[aindx]{Author, S.} % or \aindx{Vento, V.}

\address{$^1$ Department of Physics,
Chungnam National University \\ Daejon 305-764, Korea \\
bypark@cnu.ac.kr \\
$^2$ Departament de Fisica Te\`orica and Institut de F\'{\i}sica
Corpuscular \\
Universitat de Val\`encia and Consejo Superior
de Investigaciones Cient\'{\i}ficas \\
E-46100 Burjassot (Val\`encia), Spain \\
Vicente.Vento@uv.es }

\begin{abstract}
We review an approach, developed over the past few years, to describe hadronic matter
at finite density and temperature, whose underlying theoretical framework is the
Skyrme model, an effective low energy theory rooted in large $N_c$ QCD.
In this approach matter is described by various crystal structures of skyrmions,
classical topological solitons carrying baryon number, from which conventional baryons appear
by quantization.
Chiral and scale symmetries play a crucial role in the dynamics as described by pion,
dilaton and vector meson degrees of freedom.
When compressed or heated skyrmion matter describes a rich phase diagram which has
strong connections with the confinement/deconfinement phase transition.

\end{abstract}

\body

\section{Introduction}\label{intro}
An important issue at present is to understand the properties of
hadronic matter under extreme conditions, e.g., at high temperature
as in relativistic heavy-ion physics and/or at high density as in
compact stars. The phase diagram of hadronic matter turns out
richer than what has been predicted by
perturbative Quantum Chromodynamics (QCD).~\cite{Alam:2008xq}
Two approaches have been developed thus far to discuss this issue :
on the one hand, Lattice QCD which deals directly with quark and gluon degrees of freedom, and
on the other, effective field theories which are described in terms of hadronic fields.
We shall describe in here a formalism for the second approach based on the topological
soliton description of hadronic matter firstly introduced by Skyrme.~\cite{Skyrme:1961vq,Skyrme:1962vh}

Lattice QCD, the main computational tool accessible to highly
nonperturbative QCD, has provided much information on the
the finite temperature transition, such as  the value of the critical temperature, the type of
equation of state,  etc~\cite{Karsch:2007dt}. However, due to a notorious `sign problem',
lattice QCD could not be applied to study dense
matter. Only in the last few years, it has become possible to
simulate QCD with small baryon density.~\cite{Fromm:2008ab}
Chiral symmetry is a flavor symmetry of QCD which plays an essential
role in hadronic physics. At low temperatures and densities it is spontaneously
 broken leading to the existence of the pion.  Lattice studies seem to imply that chiral
 symmetry is restored in the high temperature and/or high baryon density phases and
 that it may go hand-in-hand with the confinement/deconfinement transition.
The quark condensate $\langle \bar{q} q\rangle$ of QCD is an order parameter
of this symmetry and decreases to zero when the symmetry is
restored.

The Skyrme model, is an effective low energy theory rooted in large $N_c$ QCD,~\cite{'tHooft:1973jz,Witten:1983tx}
which we have applied to dense and hot matter 
studies.~\cite{Lee:2003aq,Park:2003sd,Park:2008zg,Lee:2003eg,Lee:2003rj,Kalloniatis:2005kc,Kalloniatis:2004fx,Park:2008xm}
The model does not have explicit quark and gluon degrees of freedom,
and therefore one can not investigate the
confinement/deconfinement transition directly, but we may study
 the chiral symmetry restoration transition which occurs close by. The schemes which aim at approaching
 the phase transition from the hadronic side are labelled
`bottom up' schemes.
 The main ingredient associated with chiral symmetry is the pion, the Goldstone
boson associated with the spontaneously broken phase. The
various patterns in which the symmetry is realized in QCD will be
directly reflected in the in-medium properties of the pion and
consequently in the properties of the skyrmions made of it.

The most essential ingredients of the Skyrme model are the pions, Goldstone bosons associated with the
spontaneous breakdown of chiral symmetry.  Baryons arise as topological solitons of the meson
Lagrangian. The  pion Lagrangian can be realized non-linearly as
$U=\exp(i\vec{\tau}\cdot\vec{\pi}/f_\pi)$, which transforms as
$U \rightarrow g_L U g_R^\dagger$ under the global chiral
transformations $SU_L(N_f) \times SU_L(N_f)$; $g_L \in SU_L(N_f)$ and $g_R \in SU_R(N_f)$.
Hereafter, we will restrict our consideration to $N_f=2$.
In the case of $N_f = 2$, the meson field $\pi$ represents three pions as
\begin{equation}
\vec{\tau}\cdot\vec{\pi} =
\left(
\begin{array}{cc}
 \pi^0 & {\sqrt2}\pi^+ \\
{\sqrt2} \pi^- & -  \pi^0
\end{array}
\right).
\end{equation}
The Lagrangian for their dynamics can be expanded in powers of
the right and left invariant currents $R_\mu = U \partial_\mu U^\dagger$ and
$L_\mu = U^\dagger \partial_\mu U$, which transforms as
$R_\mu \rightarrow g_L R_\mu g_L^\dagger$ and $L_\mu \rightarrow g_R L_\mu g_R^\dagger$.
The lowest order term is
\begin{equation}
{\cal L}_{\sigma} = \frac{f_\pi^2}{4} \mbox{tr} (\partial_\mu U^\dagger \partial^\mu U).
\label{kinetic}
\end{equation}
Here, $f_\pi$ = 93 MeV is the pion decay constant.

Throughout this paper, we take the following convention
for the indices: (i) $a,b,\cdots =1,2,3$ (Euclidean metric) for
the isovector fields; (ii) $i,j,\cdots=1,2,3$ (Euclidean metric)
for the spatial components of normal vectors;  (iii)
$\mu,\nu,\cdots=0,1,2,3$ (Minkowskian metric) for the space-time
4-vectors; (iv) $\alpha,\beta,\cdots=0,1,2,3$ (Euclidean metric)
for isoscalar(0)+ isovectors(1,2,3).

In the next order, one may find three independent terms consistent with 
Lorentz invariance, parity and $G$-parity as
\begin{equation}
{\cal L}_4 = \alpha \mbox{tr} [ L_\mu, L_\nu ]^2 + \beta \mbox{tr} \{L_\mu , L_\nu \}_+
+ \gamma \mbox{tr} ( \partial_\mu L_\nu )^2.
\end{equation}
In his original work,~\cite{Skyrme:1961vq,Skyrme:1962vh}
Skyrme introduced only the first term to be denoted as
\begin{equation}
{\cal L}_{\rm sk} = \frac{1}{32 e^2}  \mbox{tr} [ L_\mu, L_\nu ]^2,
\label{skyrme}
\end{equation}
which it is still second order in the time derivatives.
The value of the ``Skyrme parameter" may be evaluated by using $\pi\pi$ data.
In the Skyrme model, it is also determined, for example,
as $e=5.45$ ~\cite{Adkins:1983ya}
to fit the nucleon-Delta masses,
or as $e = 4.75$ ~\cite{Jackson:1983bi}
to fit the axial coupling constant of nucleon.

One may build up higher order terms with more and more phenomenological
parameters. However, this naive derivative expansion leads to a Lagrangian
which has an excessive symmetry; that is, it is invariant under
$U \leftrightarrow U^\dagger$, which is not a genuine symmetry of QCD.
To break it, we need the Wess-Zumino-Witten term.~\cite{Witten:1983tw}
The corresponding action can
be written locally as
\begin{equation}
S_{WZW} = -\frac{i N_c}{240 \pi^2} \int d^5 x \varepsilon^{\mu\nu\lambda\rho\sigma}
\mbox{tr} (L_\mu L_\nu \cdots L_\sigma ),
\end{equation}
i.e. in a five dimensional space whose boundary is the
ordinary space and time. For $N_f=2$ this action vanishes trivially,
but for $N_f =3$ it provides a hypothesized process
$KK \rightarrow \pi^+ \pi^0 \pi^-$. When the action is $U(1)$ gauged for the pions to
interact with photons, this term plays a nontrivial role even with two flavors.

Chiral symmetry is explicitly broken  by the quark masses, which provides
the masses to the Goldstone bosons. The mass term can be incorporated in the same
way as chiral symmetry is broken in QCD; that is,
\begin{equation}
{\cal L}_{\rm m}
= \frac{f_\pi^2 m_\pi^2}{4} \mbox{tr} ((U+U^\dagger -2))
\sim
 - \frac{\langle \bar{q} q \rangle}{4} \mbox{tr}( {\cal M}(U + U^\dagger -2)) ,
 \label{mass}
\end{equation}
where
\begin{equation}
{\cal M} = \left( \begin{array}{cc} m_u & 0 \\ 0 & m_d \end{array} \right).
\end{equation}
We neglect the u- and d-quark mass difference.

The approach has been generalized to more sophisticated meson Lagrangians which are constructed
by implementing the symmetries of QCD.~\cite{Weinberg:1978kz}
The scale dilaton has been incorporated
into the effective scheme to describe in hadronic language
the scale anomaly.~\cite{Migdal:1982jp,Ellis:1984jv}
The vector mesons $\rho$ and $\omega$ with masses $m_{\rho, \omega} \sim 780 $ MeV
can be incorporated into the Lagrangian by using the hidden local symmetry (HLS)
~\cite{Bando:1987br} and guided by the matching of this framework
to QCD in what is called `vector manifestation' (VM).~\cite{Harada:2003jx}
We shall discuss these generalizations, when required
in the discussion of skyrmion matter, later on.

The classical nature of skyrmions enables us to construct a dense
system quite conveniently by putting more and more skyrmions into a
given volume. Then, skyrmions shape and arrange themselves to
minimize the energy of the system. The ground state configuration of
skyrmion matter are crystals. At low density it is made of
well-localized single skyrmions.~\cite{Klebanov:1985qi} At a critical
density, the system undergoes a structural phase transition to a new
kind of crystal. It is made of `half-skyrmions' which are still
well-localized but carry only  half winding number. In the
half-skyrmion phase, the system develops an additional symmetry
which leads to a vanishing average value of $\sigma=\frac{1}{2} Tr(U)$, 
the normalized trace of the $U$ field.~\cite{Goldhaber:1987pb} In the studies of the late
80's,~\cite{Jackson:1988ti} the vanishing of this average value $\langle \sigma \rangle$ was
interpreted as chiral symmetry restoration by assuming that $\langle
\sigma \rangle$ is related to the QCD order parameter $\langle
\bar{q}q\rangle$. However, in Ref. \refcite{Lee:2003aq}, it was shown that the
vanishing of $\langle \sigma \rangle$ cannot be an indication of a
genuine chiral symmetry restoration, because the decay constant of
the pion fluctuating in such a half-skyrmion matter does not vanish.
This was interpreted as a signal of the appearance of  a pseudogap phase
similar to what happens in high $T_c$ superconductors.~\cite{Tesanovic:2002zz}

The puzzle was solved in Ref. \refcite{Lee:2003eg} by incorporating a
suitable degree of freedom, the dilaton field $\chi$, associated
to the scale anomaly of QCD. The dilaton field takes
over the role of the order parameter for  chiral symmetry
restoration. As the density of skyrmion matter increases, both
$\langle \sigma\rangle$ and $\langle \chi \rangle$ vanish (not
necessarily at the same critical density). The effective decay
constant of the pion fluctuation vanishes only when $\langle
\chi\rangle$ becomes zero. It is thus the dilaton field which
provides the mechanism for chiral symmetry restoration.

Contrary to lattice QCD, there are few studies on the
temperature dependence of skyrmion matter.
Skyrmion matter has been heated up to melt the crystal into a
liquid to investigate the crystal-liquid phase transition
~\cite{Kaelbermann:1998wp,Schwindt:2002we} a phenomenon which is irrelevant to the restoration
of chiral symmetry. We have studied skyrmion matter at {\em finite density
and temperature} and have obtained the phase diagram
describing the realization of the chiral symmetry.~\cite{Park:2008xm}

The contents of this review are as follows. Section \ref{matterSK} deals with
the history of skyrmion matter and how our work follows from previous investigations.
We also study of the pion properties inside skyrmion matter
at finite density. To confront the results with reality, in Sec. \ref{dilatonSK} we
show that the scale dilaton has to be incorporated and we discuss how the properties of
the pion change thereafter. Section \ref{hotSK} is devoted to the study of the temperature dependence and the
description of the phase diagram.
In Sec. \ref{vectorSK} we incorporate vector mesons to the scheme and discuss the problem
that arises due to the coupling of the $\omega$ meson and our solution to it.
Finally the last section is devoted to a summary
of our main results and to some conclusions we can draw from our study.

\section{Matter at finite density}\label{matterSK}

\subsection{Skyrmion matter}\label{denseSK}

The Skyrme model describes baryons, with arbitrary baryon
number, as static soliton solutions of an effective Lagrangian for
pions.~\cite{Skyrme:1961vq,Skyrme:1962vh} The model has been used to describe not only single baryon
properties,~\cite{Adkins:1983ya,Zahed:1986qz} but also has served to derive the
nucleon-nucleon interaction,~\cite{Skyrme:1962vh,Jackson:1985bn} the pion-nucleon
interaction,~\cite{Schwesinger:1988af} properties of light nuclei and of
nuclear matter. In the case of nuclear matter, most of the
developments~\cite{Klebanov:1985qi,Goldhaber:1987pb,Castillejo:1989hq,Kugler:1988mu,Manton:1994rf}
 done in late 80's involve a  crystal of skyrmions.

\begin{figure}
\centerline{\epsfig{file=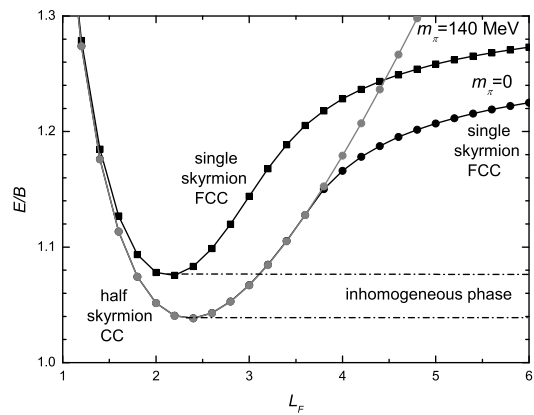,width=7cm,angle=0}}
 \caption{Energy per single skyrmion as a function of the size parameter $L$. The
solid circles show the results for massless pions and the open
circles are those for massive pions. Note the rapid phase
transition around $L\sim 3.8$ for massless pions.}\label{energy}
\end{figure}

The first attempt to understand the dense skyrmion matter was made by Kutchera {\it et
al.}~\cite{Kutschera:1984zm} These authors proceeded by introducing a single skyrmion
into a spherical Wigner-Seitz cell  without incorporating explicit information on the interaction.
The presently considered conventional approaches were developed later. In them
one assumes that the skyrmions form a crystal with a specific symmetry
and then performs numerical simulations using this symmetry as a constraint.
The first guess at this symmetry was made by Klebanov.~\cite{Klebanov:1985qi}
He considered a system where the skyrmions are located in the lattice site of a
cubic crystal (CC) and have relative orientations in such a way
that the pair of nearest neighbors attract maximally. Goldhaber
and Manton ~\cite{Goldhaber:1987pb} suggested that contrary to Klebanov's
findings, the high density phase of skyrmion matter is to be
described by a body-centered crystal (BCC) of half skyrmions. This
suggestion was confirmed by numerical calculations.~\cite{Jackson:1988ti}
Kugler and Shtrikman,~\cite{Kugler:1989uc} using a variational method,
investigated the ground state of the skyrmion crystal including
not only the single skyrmion CC and half-skyrmion BCC but also the
single skyrmion face-centered-cubic crystal (FCC) and half-skyrmion CC.
In their calculation a phase transition from the single FCC  to half-skyrmion CC
takes place and the ground state is found in the half-skyrmion CC configuration.
Castillejo {\it et al. } ~\cite{Castillejo:1989hq} obtained similar conclusions.

In Fig. \ref{energy} we show the energy per baryon $E/B$ as a function of the
FCC box size parameter $L$ \footnote{A single FCC is a cube with a sidelength $2L$, so that there are
4 single skyrmions in a volume of $8L^3$, that is, the baryon number density is related to $L$ as
$\rho_B = 1/2L^3$.}. Each point in the figure denotes a minimum of the energy
for the classical field configuration associated with the Lagrangians
(\ref{kinetic}), (\ref{skyrme}) and (\ref{mass}) for a given value of $L$.
The solid circles correspond to the zero pion mass calculation and reproduce the results of Kugler
and Shtrikman.~\cite{Kugler:1988mu} The quantities
$L$ and $E/B$, appearing in the figure,  are given in units of $(ef_\pi)^{-1}$ $(\sim 0.45$fm
with $f_\pi=93$ MeV and $e=4.75$) and $E/B$ in units of $(6\pi^2 f_\pi)/e$ $(\sim
1160$ MeV), respectively. The latter enable us to compare the
numerical results on $E/B$ easily with its Bogolmoln'y bound for
the skyrmion in the chiral limit, which can be expressed as
$E/B=1$ in this convention.

In the chiral limit, as we squeeze the system
from $L=6$ to around $L =3.8$, one sees that the skyrmion system undergoes a
phase transition from the FCC single skyrmion configuration to the
CC half-skyrmion configuration. The system reaches a minimum energy
configuration at $L=L_{\rm min}\sim 2.4$ with the energy per baryon $E/B\sim 1.038$.
This minimum value is  close to the Bogolmoln'y bound for the energy associated to
Eqs.(\ref{kinetic}) and (\ref{skyrme}).

On the other hand, the configuration  found at $L>L_{\rm min}$ with
the constrained symmetry may not be the genuine low energy configuration of the system
for that given $L$. Note that the pressure $P\equiv\partial E/\partial V$ is negative,
which implies that the system in that configuration is unstable. Some of the skyrmions
may condense to form dense lumps in the phase
leaving large empty spaces forming a stable inhomogeneous as seen in Fig. \ref{energy}  for $L=L_{\rm min}$.
Only the phase to the left of the minimum, $L<L_{\rm min}$, may be referred to
as ``{\em homogeneous}" and there the background field is
described by a crystal configuration.

\begin{figure}
\centerline{\psfig{file=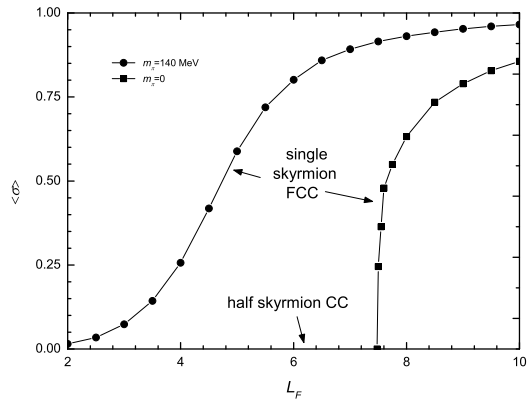,width=7cm,angle=0}}
\caption{$\langle \sigma\rangle$ as a function of the size
parameter $L$. The notation is the same as in Fig. \ref{energy}}
\label{sigmask}
\end{figure}

The open circles are the solutions found with a nonvanishing pion
mass, $m_\pi=140$ MeV.\footnote{Incorporating the pion mass into
the problem introduces a new scale in the analysis and therefore
we are forced to give specific values to the parameters of the
chiral effective Lagrangian, the pion decay constant and the
Skyrme parameter, a feature which we have avoided in the chiral
limit. In order to proceed, we simply take their empirical values,
that is, $f_\pi=93$ MeV and $e=4.75$. Although the numerical
results depend on these values, their qualitative behavior will
not.} Comparing to the skyrmion system for massless pions, the
energy per baryon is somewhat higher. Furthermore, there is no
first order phase transition at low densities.

\begin{figure}
\centerline{\psfig{file=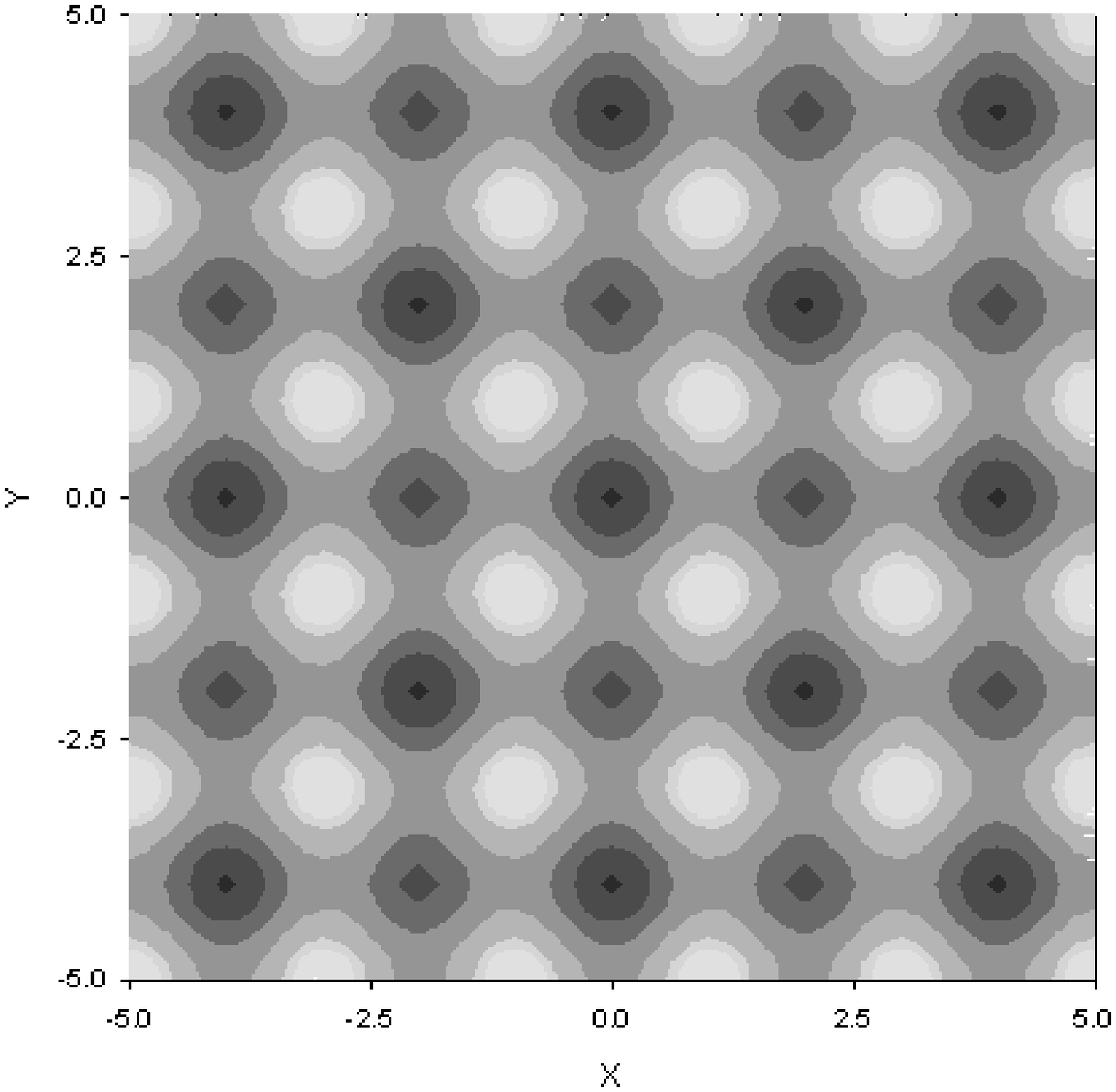,width=6cm,angle=0}
\psfig{file=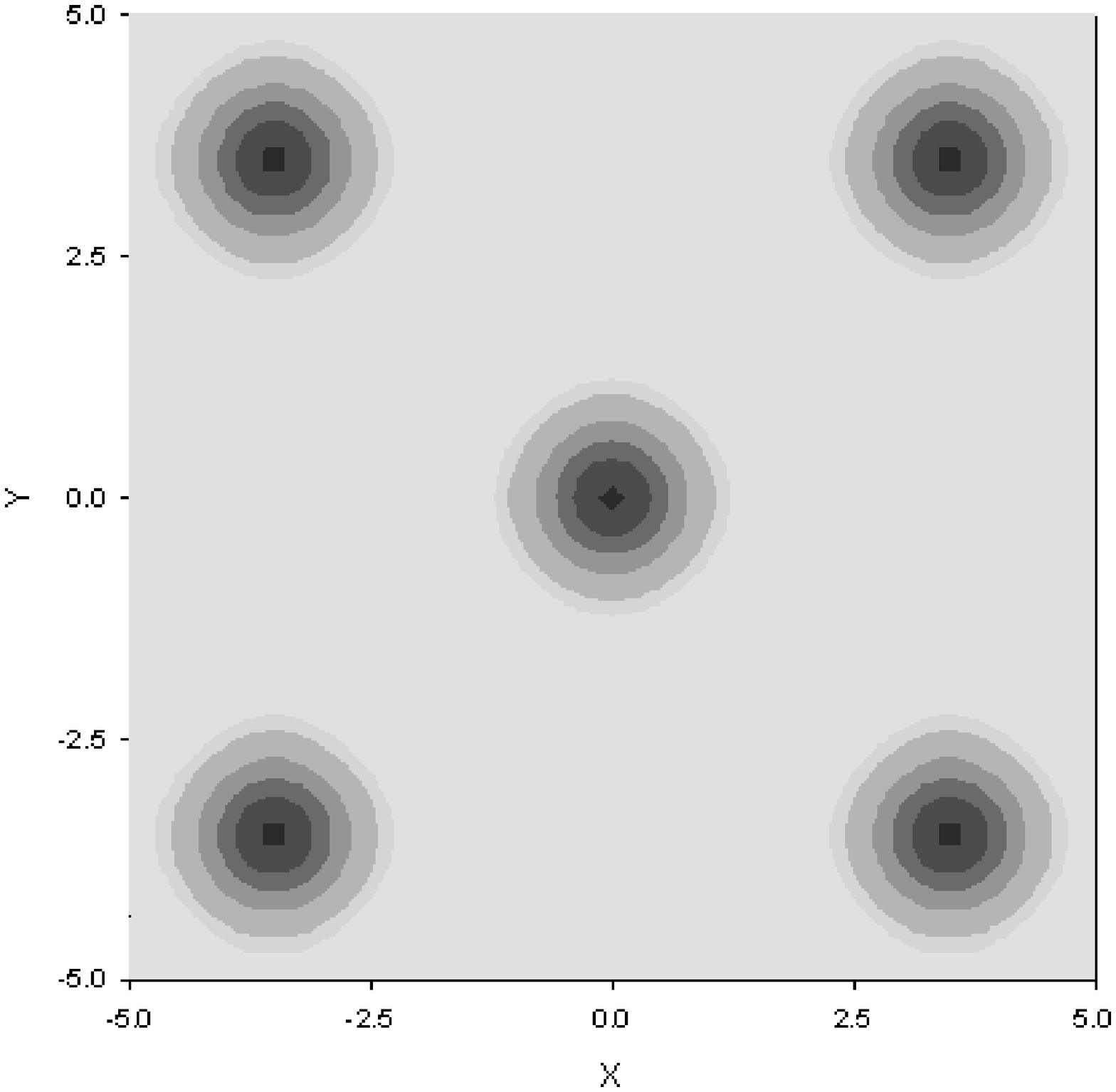,width=6cm,angle=0}}
 \caption{Local baryon
number densities at $L=3.5$ and $L=2.0$ with massive pions. For
$L=2.0$ the system is (almost) a half-skyrmion in a CC crystal
configuration.}
\label{density}
\end{figure}

In Fig. \ref{sigmask}, we show $\langle \sigma \rangle$, 
i.e. the space average value of $\sigma$ as a function of $L$.
In the chiral limit, $\langle \sigma\rangle$ rapidly drops as the system
shrinks and reaches  zero at $L\sim 3.8$, where the system goes to
the half-skyrmion phase. This phase transition was interpreted ~\cite{Forkel:1989wc}
as a signal for chiral symmetry restoration.
However, as we sall see in the next section, this is not the expected transition.
In the case of massive pions, the
transition in $\langle \sigma\rangle$ is soft. Its value
 decreases monotonically and reaches zero asymptotically, as the
density increases. Furthermore, as we can see in Fig. \ref{density}, where the
local baryon number density is shown, for $L=2$ (left) and $L=3.5$
(right) in the $z=0$ plane, the system becomes a half-skyrmion
crystal at high density.

\begin{figure}
\centerline{\psfig{file=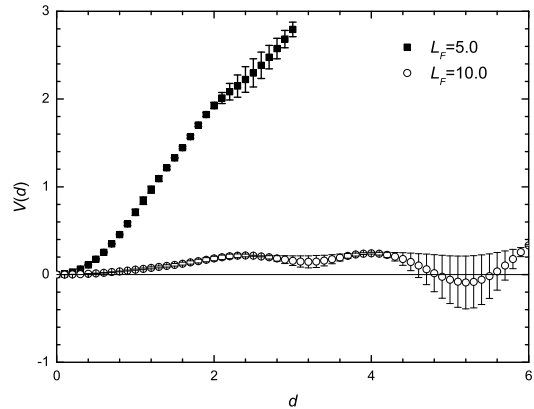,width=7cm,angle=0}}
\caption{The energy cost to shift a single skyrmion from its stable position
by an amount $d$ in the direction of the $z$-axis.}
\label{potE}
\end{figure}

Another scheme used to study multi-skyrmion systems
is the procedure based on the Atiyah-Manton Ansatz.~\cite{Atiyah:1989dq}
In this scheme, skyrmions of baryon number $N$ are obtained by
calculating the holonomy of Yang-Mills instantons of charge $N$.
This Ansatz has been  successful in describing few-nucleon systems.
~\cite{Atiyah:1992if,Leese:1993mc,Walet:1996rc}  This procedure has been also
applied to nuclear matter with the instanton solution on a four
torus.~\cite{Manton:1994rf} The energy per baryon was found to be
$(E/B)_{min} = 1.058$ at $L_{\min}= 2.47$, which is comparable to
the variational result of Kugler and Shtrikman.~\cite{Kugler:1989uc}

In Ref. \refcite{Park:2002ie}, the Atiyah-Maton Ansatz is employed to
get skyrmion matter from the 't Hooft's multi-instanton solution, which
is modified to incorporate dynamical variables such as the positions
and relative orientations of the single skyrmions. This description provides
information on the dynamics of a single skyrmion in skyrmion matter.
Shown in Fig. \ref{potE} is the energy change of the system when a single
skyrmion is shifted from its FCC lattice site by an amount $d$
in the direction of the z-axis. Two extreme cases are shown.
In the case of a dense system ($L_F \equiv 2L = 5.0$), the energy changes abruptly.
For small $d$, it is almost quadratic in $d$. It implies that the dense system
is in a crystal phase. On the other hand, in the case of a dilute system
($L_F = 10.0$), the system energy remains almost constant up to some large $d$,
which implies that the system is in a gas (or liquid) phase. If we let all
the variables vary freely, the system will prefer to change to a disordered
or inhomogeneous phase in which some skyrmions will form clusters,
as we have discussed before.

\subsection{Pions in Skyrmion matter} \label{pionSK}

The Skyrme model also provides the most convenient framework to study the pion properties
in dense matter. The basic strategy is to take the static configuration
$U_0(\vec{x})$ discussed in Sec. \ref{denseSK}
as the background fields and to look into the properties of the pion fluctuating on top of it.
This is the conventional procedure used to find single particle excitations when one has solitons
in a field theory.~\cite{Jackiw:1977rj}

The fluctuating {\em time-dependent} pion fields can be incorporated on top of the
static fields through the Ansatz~\cite{Saito:1986oy}
\begin{equation}
U(\vec{x},t)=\sqrt{U_\pi } U_0 (\vec{x})\sqrt{U_\pi},
\end{equation}
where
\begin{equation}
U_\pi=\exp\bigg(i\vec{\tau}\cdot\vec{\phi}(x)/f_\pi\bigg),
\end{equation}
with $\vec{\phi}$ describing the fluctuating pions.

When $U_0(\vec{r})=1(\rho_B=0)$, the expansion in power of $\phi$'s leads us to
\begin{equation}
{\cal L}(\phi) = \frac12 \partial_\mu \phi^a \partial^\mu \phi^a
 + \frac12 m_\pi^2 \sigma(\vec{x}) \phi^a \phi^a
 + \cdots,
\end{equation}
which is just a Lagrangian for the self-interacting pion fields
without any interactions with baryons.
Here, we have written only the kinetic  and mass terms relavant for further discussions.
With a non trivial $U_0(\vec{r})$ describing  dense skyrmion matter,  the Lagrangian becomes,
\begin{equation}
{\cal L}= \frac12 G^{ab}(\vec{x}) \partial_\mu \phi_a \partial^\mu \phi_b
  + \frac12 m_\pi^2 \sigma(\vec{x}) \phi^a \phi^a
 + \cdots,
\end{equation}
with
\begin{equation}
G^{ab}(\vec{x}) = \sigma^2 \delta_{ab} + \pi_a \pi_b.
\label{G}
\end{equation}

The structure of our Lagrangian is similar to
that of chiral perturbation theory Eq. (\ref{ChPT}) of
Refs.~\refcite{Yabu:1994de,Thorsson:1995rj}. These authors start with a Lagrangian containing
all the degrees of freedom, including nucleon fields, and free parameters. They integrate out
the nucleons in and out of an \`a priori assumed Fermi sea and in
the process they get a Lagrangian density describing the pion in
the medium. Their result corresponds to the above Skyrme Lagrangian except
that the quadratic (current algebra) and the mass terms pick up a
density dependence of the form
\begin{equation}
-\frac{f_\pi^2}{4} \left( g^{\mu\nu} +  \frac{D^{\mu\nu} \rho}
{f_\pi^2}\right) \mbox{Tr} (U^\dagger \partial_\mu U U^\dagger
\partial^\nu U)
 + \frac{f_\pi^2 m_\pi^2}{4}
\left( 1 - \frac{\Sigma_{\pi N}}{f_\pi^2 m_\pi^2}  \rho \right)
\mbox{Tr} (U+U^\dagger -2 ), \label{ChPT}\end{equation}
where $\rho$ is the density of the nuclear matter and $D^{\mu\nu}$
and $\sigma$ are physical quantities obtained from the
pion-nucleon interactions. Note that in this scheme, nuclear
matter is assumed ab initio to be a Fermi sea devoid of the
intrinsic dependence mentioned above.

\begin{figure}
\centerline{\epsfig{file=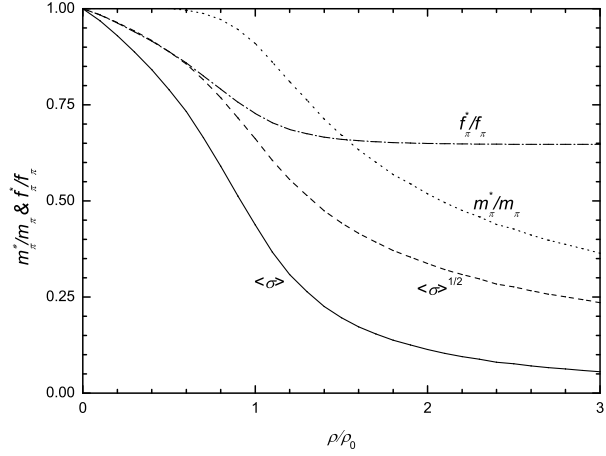,width=8cm,angle=0}}
\caption{Estimates of $f^*_\pi/f_\pi$ and $m^*_\pi/m_\pi$ as
functions of the baryon number density of skyrmion matter.}
\label{mpistar}
\end{figure}

We proceed via a mean field approximation  consisting in
averaging the background modifications $G^{ab}(\vec{x})$ and
$\sigma(\vec{x})$ appearing in the Lagrangian which
are reduced to constants, $\langle G^{ab} \rangle = G \delta_{ab}$
and $\langle \sigma \rangle$. Then, the Lagrangian can be rewritten as
\begin{equation}
{\cal L}(\phi^*) = \frac12 \partial_\mu \phi^*_a \partial^\mu \phi^*_a
+ \frac12 m_\pi^* \phi^*_a \phi^*_a + \cdots ,
\end{equation}
where we have carried out a wavefunction renormalization,
$\phi^*_a = \sqrt{G} \phi_a$, which leads to a medium modified pion decay
constant and mass as
\begin{eqnarray}
\frac{f_\pi^*}{f_\pi} = \sqrt{G}, \\
\frac{m_\pi^*}{m_\pi} = \frac{\langle \sigma \rangle}{\sqrt G}.
\end{eqnarray}

In Fig. \ref{mpistar} we show the estimates of $f^*_\pi/f_\pi$ and
$m^*_\pi/m_\pi$ as a function of the density. As the density
increases, $f^*_\pi$ decreases only to $\sim 0.65 f_\pi$ and then it
remains constant at that value.
Our result is different from what was the general believe:~\cite{Forkel:1989wc}
{\em the vanishing of
$\langle \sigma \rangle$ is not an indication of chiral symmetry restoration since
 the pion decay constant does not vanish.}

Note that $\langle \sigma\rangle^\frac12$ has the
same slope at low densities, which leads to $m^*_\pi/m_\pi \sim 1$
at low densities. Since at higher densities $G$
becomes a constant, $m^*_\pi/m_\pi$ decreases like $\langle
\sigma\rangle^{1/2}$ with a factor which is greater than 1. As the
density increases, higher order terms in $\rho$ come to play
important roles and $m^*_\pi/m_\pi$ decreases.
A more rigorous derivation of these quantities can be obtained
using perturbation theory.~\cite{Lee:2003aq}

The slope of $\langle \sigma\rangle$ at low density is
approximately 1/3. If we expand $\langle\sigma\rangle$ about
$\rho=0$ and compare it with Eq.(\ref{ChPT}), we obtain
\begin{equation}
\langle\sigma\rangle \sim 1 - \frac{1}{3} \frac{\rho}{\rho_0} +
\cdots \sim 1 -\frac{\Sigma_{\pi N}}{f_\pi^2 m_\pi^2} \rho +
\cdots,
\end{equation}
which yields $\Sigma_{\pi N} \sim m^2_\pi f_\pi^2 /(3\rho_0) \sim
42$ MeV, which is comparable with the experimental value 45
MeV~\footnote{While this value is widely quoted, there is a considerable
controversy on the precise value of this
sigma term. In fact it can even be considerably higher than this.
See Ref. \refcite{Gibbs:2003ka} for a more recent discussion.}. This comparison is
fully justified from the point of view of the $\frac{1}{N}$
expansion since both approaches should produce the same result to
leading order in this expansion. The liner term is ${\cal{O}}$(1).

The length scale is strongly dependent on our choice of the
parameters $f_\pi$ and $e$. Thus one should be aware that the
$\rho$ scale  in Fig. \ref{mpistar} could change quantitatively considerably if
one chooses another parameter set, however the qualitative
behavior will remain unchanged.

 Note that the density
dependence of the background is taken into
account to all orders. No low-density approximation, whose validity
is in doubt except at very low density, is ever made in the
calculation. The power of our approach is that the dynamics of the
background and its excitations  can be treated in a unified way
on the same footing with a single Lagrangian.

\section{Implementing scale invariance}\label{dilatonSK}

\subsection{Dilaton dynamics}\label{DD}
The dynamics introduced in Sec. \ref{intro} as an effective theory for the hadronic interactions
 is probably incomplete. In fact, it is not clear
that the intrinsic density dependence required by the matching to
QCD is fully implemented in the model. One puzzling feature is that the Wigner
phase represented by the half-skyrmion matter with
$\langle  \sigma \rangle=0$ supports a non-vanishing pion decay constant.
This may be interpreted as a possible signal for a pseudogap phase.
 However, at some point, the chiral symmetry should be restored
and there the pion decay constant should vanish.

This difficulty can be circumvented in our
framework by incorporating in the standard skyrmion dynamics the trace
anomaly of QCD in an effective manner.~\cite{Migdal:1982jp}
The end result is the skyrmion Lagrangian
introduced by Ellis and Lanik~\cite{Ellis:1984jv} and employed by Brown
and Rho~\cite{Brown:1991kk} for nuclear physics which contains an
 additional scalar field, the so called scale
dilaton.

The classical QCD action of scale dimension 4 in the chiral limit
is invariant under the scale transformation
\begin{equation}
x \rightarrow {}^\lambda x = \lambda^{-1} x, \ \ \ \lambda \geq 0,
\end{equation}
under which the quark field and the gluon fields transform with
the scale dimension 3/2 and 1, respectively. The quark mass term
of scale dimension 3 breaks scale invariance. At the quantum
level, scale invariance is also broken by dimensional
transmutation even for massless quarks, as signaled by the
non-vanishing of the trace of the energy-momentum tensor. Equivalently,
this phenomenon  can be formulated by the non-vanishing divergence
of the dilatation current $D_\mu$, the so called trace anomaly,
\begin{equation}
\partial^\mu D_\mu = \theta^\mu_\mu
= \sum_q m_q \bar{q}q - \frac{\beta(g)}{g} \mbox{Tr}G_{\mu\nu} G^{\mu\nu},
\label{TraceAnomaly}
\end{equation}
where $\beta(g)$ is the beta function of QCD.

Broken scale invariance can be implemented into  large $N_c$
physics by modifying the standard skyrmion Lagrangian, introduced
in Sec. \ref{intro}, to
\begin{eqnarray}
{\cal L}
&=& \frac{f_\pi^2}{4} \left(\frac{\chi}{f_\chi}\right)^2
{\rm Tr} (\partial_\mu U^\dagger \partial^\mu U)
+\frac{1}{32e^2} {\rm Tr}
([U^\dagger\partial_\mu U, U^\dagger\partial_\nu U])^2 \nonumber\\
&&+\frac{f_\pi^2m_\pi^2}{4} \left(\frac{\chi}{f_\chi}\right)^3
{\rm Tr} (U+U^\dagger-2) \nonumber \\
&&  +\frac{1}{2}\partial_\mu \chi\partial^\mu \chi
-\frac{1}{4}\frac{m_\chi^2}{f_\chi^2}\bigg[\chi^4\bigg(\ln(\chi/f_\chi)
-\frac{1}{4}\bigg)+\frac{1}{4}\bigg].
 \label{lag-chi}
\end{eqnarray}
We have denoted the non vanishing vacuum expectation value of
$\chi$ as $f_\chi$, a constant which describes
the decay of the scalar into pions. The second term of the trace
anomaly (\ref{TraceAnomaly}) can be reproduced by the potential
energy $V(\chi)$, which is adjusted in the
Lagrangian~(\ref{lag-chi}) so that $V=dV/d\chi=0$ and $d^2
V/d\chi^2=m_\chi^2$ at $\chi=f_\chi$.~\cite{Migdal:1982jp}

The vacuum state of the Lagrangian at zero baryon number density
is defined by $U=1$ and $\chi= f_\chi$.
The fluctuations of the pion and the scalar
fields about this vacuum, defined through
\begin{equation}
U=\exp(i\vec{\tau}\cdot\vec{\phi}/f_\pi),
\mbox{\ \ and \ \ }
\chi = f_\chi + \tilde{\chi}
\end{equation}
give physical meaning to the model parameters: $f_\pi$ as the pion
decay constant, $m_\pi$ as the pion mass, $f_\chi$ as the scalar
decay constant, and $m_\chi$ as the scalar mass. For the pions, we
use their empirical values as $f_\pi=93$MeV and $m_\pi=140$MeV. We
fix the Skyrme parameter $e$ to 4.75 from the axial-vector
coupling constant $g_A$ as in Ref. \refcite{Brown:1984sx}. However, for the
scalar field $\chi$, no experimental values for the corresponding
parameters are available.

In Ref. \refcite{Furnstahl:1995by}, the scalar field is incorporated into a
relativistic hadronic model for nuclear matter not only to account
for the anomalous scaling behavior but also to provide the
mid-range nucleon-nucleon attraction. Then, the parameters
$f_\chi$ and $m_\chi$ are adjusted so that the model fits finite
nuclei. One of the parameter sets is $m_\chi=550$ MeV and
$f_\chi=240$ MeV~(Set A). On the other hand, Song {\em et
al.}~\cite{Song:1997kx} obtain the ``best" values for the parameters of
the effective chiral Lagrangian with the ``soft" scalar fields so
that the results are consistent with  ``Brown-Rho"
scaling,~\cite{Brown:1991kk} explicitly, $m_\chi=720$ MeV and $f_\chi=240$
MeV~(Set B). For completeness, we consider also a parameter set of $m_\chi=1$
GeV and $f_\chi=240$ MeV~(Set C) corresponding to a mass scale
comparable to that of chiral symmetry $\Lambda_\chi\sim 4\pi
f_\pi$.

\subsection{Dynamics of the single skyrmion}

The procedure one has to follow can be found in Ref. \refcite{Lee:2003eg} and is similar to the one discussed in Sec. \ref{denseSK}.
The first step is to find the solution for the single skyrmion which includes the dilaton dynamics.
The skyrmion with the baryon number $B=1$ can be found
by generalizing the spherical hedgehog Ansatz of the original Skyrme
model as
\begin{equation}
U_0(\vec{r})=\exp(i\vec{\tau}\cdot\hat{r} F(r)),
\mbox{ and }
\chi_0(\vec{r}) = f_\chi C(r),
\end{equation}
with two radial functions $F(r)$ and $C(r)$.
Minimization of the mass equation leads to a coupled set
of equations of motion for these functions.
In order for the solution to carry a baryon number, $U_0$ has the
value $-1$ at the origin, that is,  $F(x=0)=\pi$, while there is no
such topological constraint for $C(x=0)$. All that is required is
that it be a positive number below 1. At infinity, the fields $U_0(\vec{r})$ and $\chi_0(\vec{r})$
should reach their vacuum values.

\begin{figure}
 \centerline{\epsfig{file=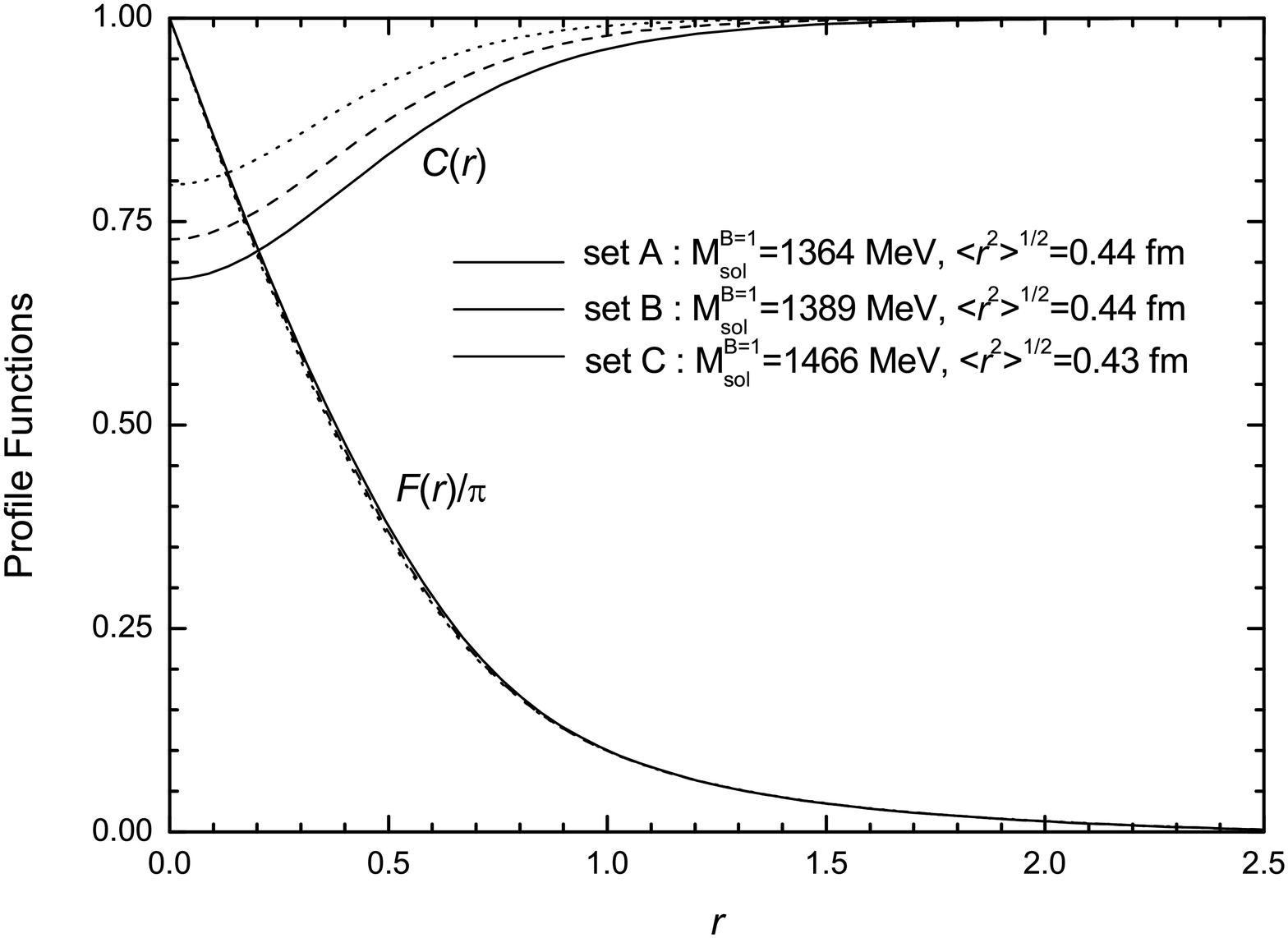,width=8cm,angle=0}}
\caption{Profile functions $F(x)$ and $C(x)$ as a function of
$x$.}
 \label{prof}
\end{figure}

Shown in Fig. \ref{prof} are profile functions as a function of
$x(=ef_\pi r)$. $F(r)$ and consequently the root mean square radius of the
baryon charge show little dependence on $m_\chi$.
On the other hand, the changes in $C(r)$ and the skyrmion mass are recognizable.
Inside the skyrmion, especially at the center, $C(r)$ deviates from its vaccum value 1.
Note that this changes in $C(r)$ is multiplied by
$f_\pi^2$ in the current algebra term of the Lagrangian. Thus,
$C(r)\leq 1$ reduces the {\em effective} $f_\pi$ inside the single
skyrmion, which implies a partial restoration of the chiral symmetry there.
The reduction in the effective pion decay constant is reflected in the single skyrmion mass.

The larger the scalar mass is, the smaller its coupling to the pionic field and
the less its effect on the single skyrmion. In the limit of
$m_\chi \rightarrow \infty$, the scalar field is completely
decoupled from the pions and the model returns back to the
original one, where $C(r)=1$, $M_{sk}=1479$ MeV and $\langle
r^2\rangle^{1/2}=0.43$ fm.

\subsection{Dense skyrmion matter and chiral symmetry restoration}\label{denseSK_dilaton}

The second step is to construct a crystal configuration made up of skyrmions
with a minimal energy for a given density.

Referring to Refs.~\refcite{Lee:2003eg,Lee:2003rj} for the full details, we emphasize here
the role the dilaton field in the phase transition scenario for skyrmion matter.
Let the dilaton field $\chi(\vec{r})$ be a constant throughout the whole space as
\begin{equation}
\chi/f_\chi = X.
\end{equation}
Then the energy per baryon number of the system for a given density can be
calculated and conveniently expressed as ~\cite{Lee:2003eg}
\begin{equation}
E/B(X,L) =  X^2 (E_2/B) + (E_4/B) + X^3 (E_m/B) +
(2L^3) \left(X^4 (\mbox{ln}X - \textstyle \frac14) + \frac14\right),
\label{E_over_B}
\end{equation}
where $E_2$, $E_4$ and $E_m$ are, respectively, the contributions
from the current algebra term, the Skyrme term and the pion mass
term of the Lagrangian to the energy of the skyrmion system, described in Sec. \ref{intro}, and
$(2L^3)$ is the volume occupied by a single skyrmion

\begin{figure}
\centerline{\epsfig{file=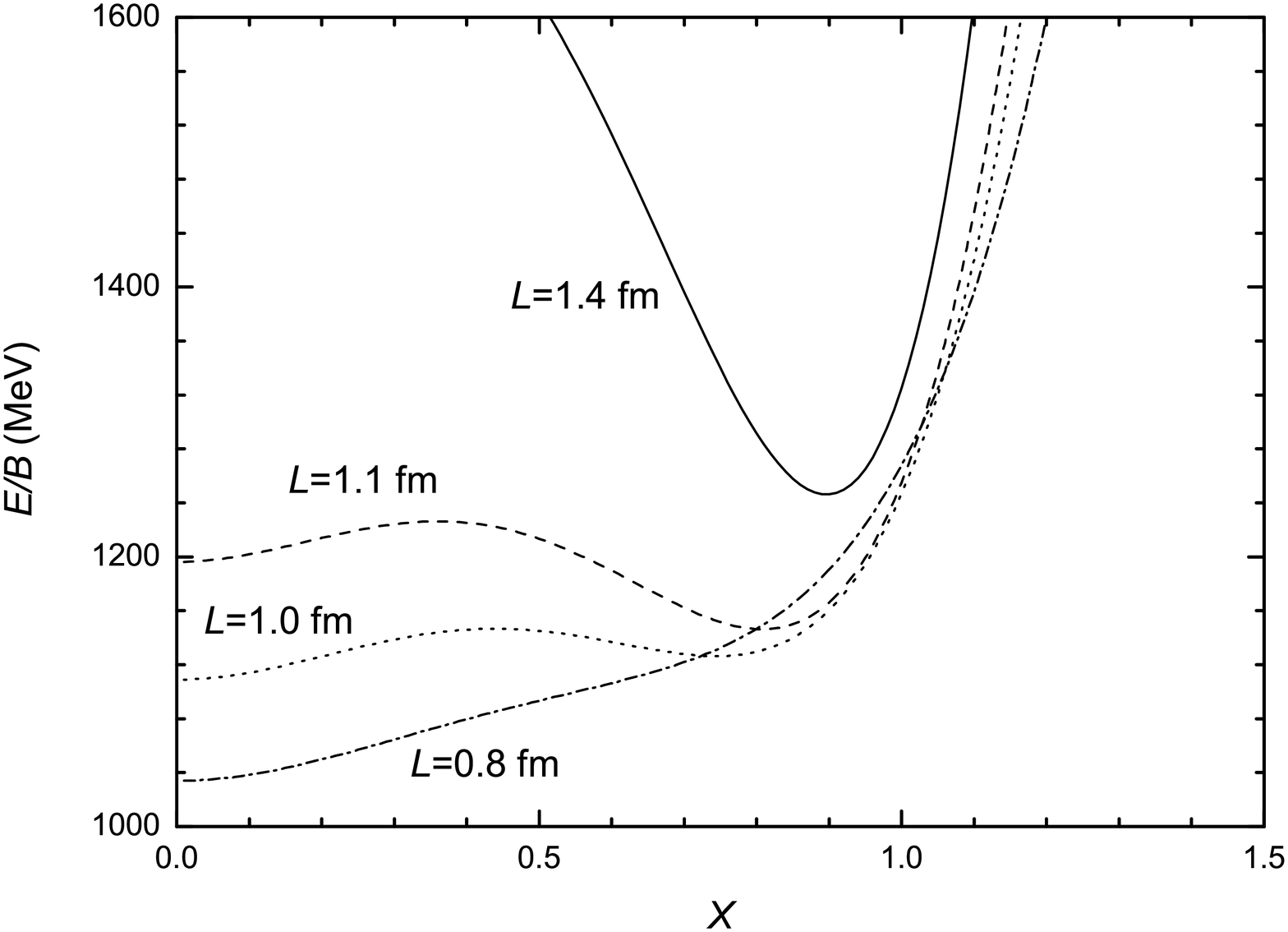,width=6cm,angle=0}
\epsfig{file=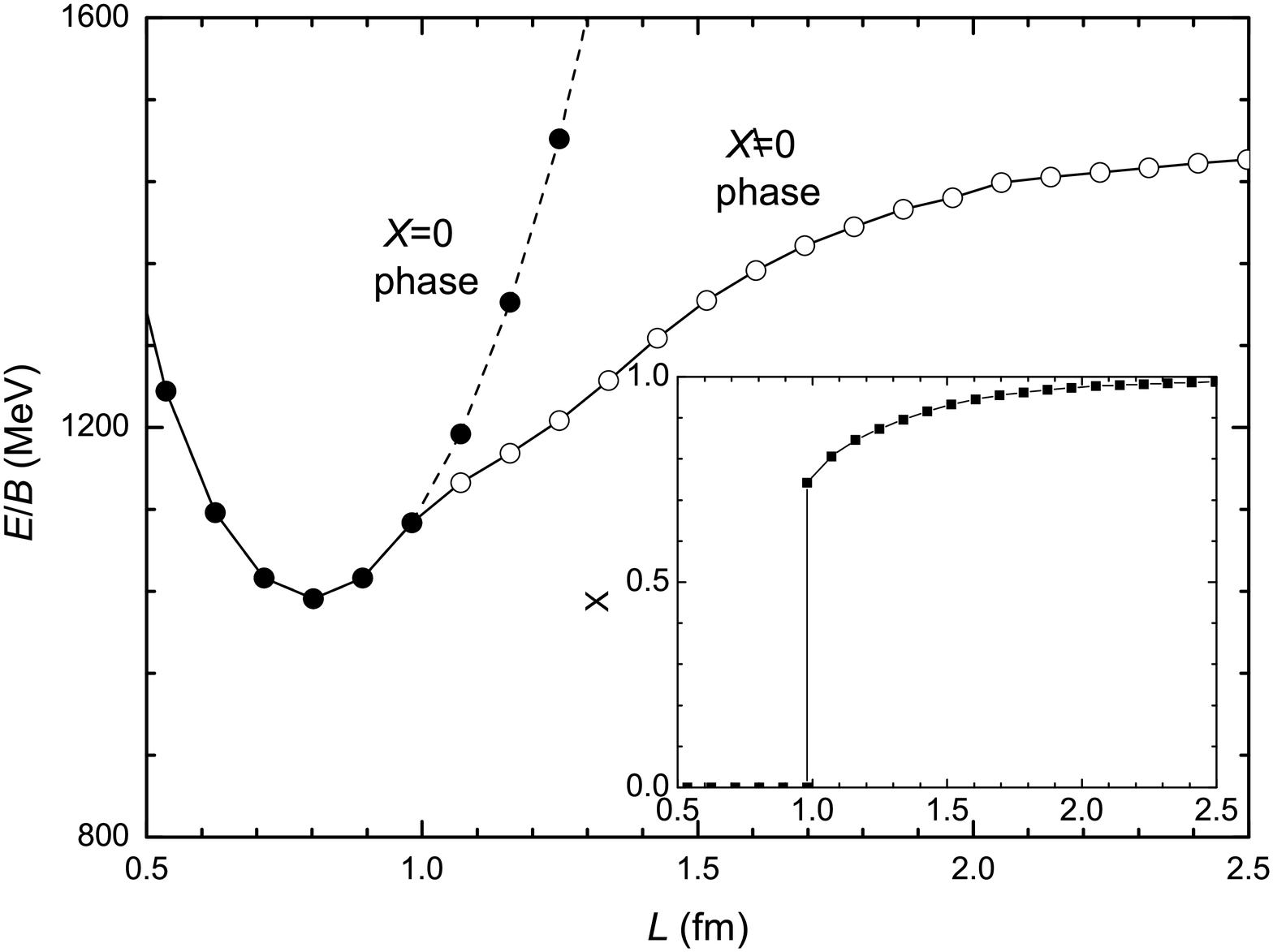,width=6cm,angle=0}}
\centerline{(a) \hskip 6cm (b)}
\caption{(a) Energy per single skyrmion as a function of the scalar
field $X$ for a given $L$. The results are obtained with the
$(E_2/B)$, $(E_4/B)$, and $(E_m/B)$ of Ref. \refcite{Lee:2003rj} and with
the parameter sets B, (b) Energy per single skyrmion as a function of $L$. }
\label{Toy}
\end{figure}

The quantity $E/B(X,L)$ can be taken as an  {\em in
medium} effective potential for $X$, modified by the coupling of the scalar to the background
matter. Using the parameter values of Ref. \refcite{Lee:2003rj}
for the Skyrme model without the scalar field, the effective
potential $E/B(X)$ for a few values of $L$ behaves as shown in Fig. \ref{Toy}(a).
At low density (large $L$), the minimum of the effective potential is
located close to $X=1$. As the density increases, the quadratic term
in the effective potential $E/B(X)$ develops another minimum at $X=0$
which is an unstable extremum of the potential $V(X)$ in free space.
At $L\sim 1$ fm, the newly developed minimum  competes with the one
near $X \sim 1$. At higher density, the  minimum shifts to
$X=0$ where the system  stabilizes.

In Fig. \ref{Toy}(b), we plot $E/B(X_{min},L)$ as a function of
$L$, which is obtained by minimizing $E/B(X,L)$ with respect to $X$
for each $L$. The figure in the small box is the corresponding value of
$X_{min}$ as function of $L$. There we see the explicit
manifestation of a first-order phase transition. Although the present
discussion is based on a simplified analysis, it essentially encodes the
same physics as in the more rigorous treatment of $\chi$ given
in Ref. \refcite{Lee:2003eg}

\begin{figure}
\centerline{\epsfig{file=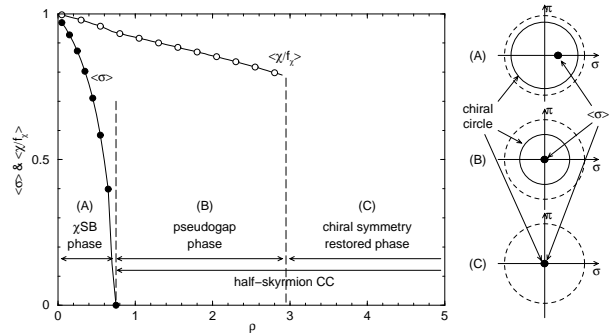,height=8cm,angle=270}}
\caption{Average values of $\sigma=\frac12\mbox{Tr}(U)$ and $\chi/f_\chi$
of the lowest energy crystal configuration at a given baryon number
density.}
\label{pseudogap}
\end{figure}

We show in Fig. \ref{pseudogap} the average values $\langle \sigma \rangle$
and $\langle \chi/f_\chi \rangle$ over space for the minimum energy crystal
configurations obtained by the  complete numerical calculation without any approximation for
 $\chi$. These data show that  a `structural' phase transition takes place, characterized by
$\langle \sigma\rangle=0$,  at lower density then  the
{\em genuine} chiral phase transition which occurs when $\langle \chi \rangle=0$.
The value of $\langle \sigma \rangle$ becomes $0$ when the structure of the skyrmion crystal
undergoes a change from the single skyrmion FCC to the half-skyrmion CC.
Thus, the  pseudogap phase persists in an intermediate
density region, where the $\langle \chi/f_\chi\rangle$ does not
vanish while $\langle \sigma \rangle$ does.~\cite{Reinhardt:1988zi}
A similar pseudogap structure has been  also proposed in hot QCD.~\cite{Zarembo:2001wr}

The two step phase transition is schematically illustrated in
Fig. \ref{pseudogap}. Let $\rho_p$ and $\rho_c$ be the density at which
$\langle \sigma\rangle$ and $\langle \chi \rangle$ vanish, repectively.
\begin{enumerate}
\item [(A)]
At low density ($\rho < \rho_p$), matter slightly reduces the
vacuum value of the dilaton field from that of the baryon free
vacuum. This implies a shrinking of the radius of the chiral
circle by the same ratio. Since the skyrmion takes all the values
on the chiral circle, the expectation value of $\sigma$ is not
located on the circle but inside the circle. Skyrmion matter at
this density is in the chiral symmetry broken phase.

\item [(B)]
At some intermediate densities ($\rho_p < \rho < \rho_c$), the
expectation value of $\sigma$ vanishes while that of the dilaton
field is still nonzero. The skyrmion crystal is in a CC
configuration made of half skyrmions localized at the points where
$\sigma=\pm 1$. Since the average value of the dilaton field does
not vanish, the radius of the chiral circle is still finite. Here,
$\langle \sigma \rangle =0$ does not mean that chiral symmetry is
completely restored. We interpret this as a pseudogap phase.

\item [(C)]
At higher density ($\rho > \rho_c$), the phase characterized by
$\langle \chi/f_\chi \rangle=0$ becomes energetically favorable.
Then, the chiral circle, describing the fluctuating pion dynamics,
shrinks to a point.

\end{enumerate}
The density range for the ocurrence of a pseudogap phase strongly depends on the parameter choice of $m_\chi$. For
small $m_\chi $ below 700 MeV, the pseudogap has almost zero size.

In the case of massive pions, the chiral circle is tilted by the
explicit (mass) symmetry breaking term. Thus, the exact half-skyrmion CC,
which requires a symmetric solution for points with value $\sigma=+1$
and those with $\sigma=-1$ cannot be constructed and consequently the phase
characterized by $\langle \sigma \rangle=0$  does not exist for any
density. Thus no pseudogap phase arises. However, $\langle \sigma \rangle$ is always inside the
chiral circle and its value drops much faster than that of
$\langle \chi/f_\chi\rangle$. Therefore, only if the pion mass is small
 a pseudogap phase can appear in the model.

\subsection{Pions in a dense medium with dilaton dynamics}

Since we have achieved, via dilaton dynamics, a reasonable scenario for chiral symmetry restoration,
it is time to revisit the properties of pions in a dense medium.
As was explained in Sec. \ref{pionSK} and in Ref. \refcite{Lee:2003aq},
we proceed to incorporate the fluctuations on top of the static skyrmion
crystal. (We refer to Refs.~ \refcite{Lee:2003eg,Lee:2003rj} for  details.)

Using a mean field approximation we calculate
the in-medium pion~ mass $m_\pi^*$ and decay constant $f_\pi^*$  obtaining,
\begin{eqnarray}
Z_\pi^2 &=& \left\langle \left(\frac{\chi_0(\vec{x})}{f_\chi} \right)^2
\textstyle (1 - \frac23 \pi^2(\vec{x})) \right\rangle
\equiv \left( \frac{f_\pi^*}{f_\pi} \right)^2,
\label{f*/f} \\
m_\pi^{*2} Z_\pi^2 &=&
\left\langle \left(\frac{\chi_0(\vec{x})}{f_\chi} \right)^3
\sigma(\vec{x})\ {m_\pi^2} \right\rangle.
\label{mp*/mp}
\end{eqnarray}
The wave function renormalization constant $Z_\pi$ gives the ratio
of the in-medium pion decay constant $f_\pi^*$ to the free one, and
the above expression arises from the current algebra term
in the Lagrangian. The explicit calculation of $m_\chi^*$ is given in Ref. \refcite{Lee:2003rj}.

\begin{figure}
\centerline{\epsfig{file=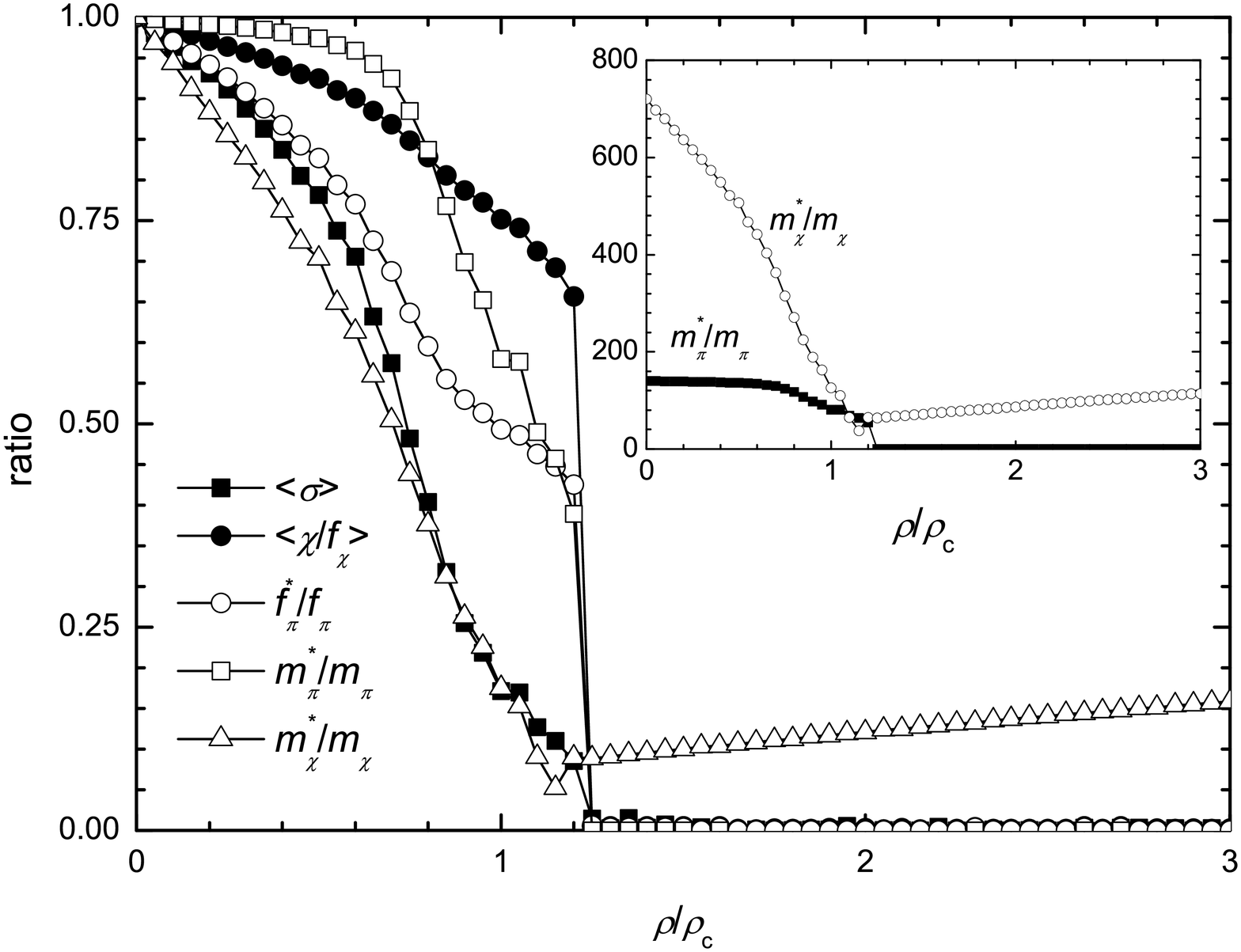,width=6cm,angle=0}
\epsfig{file=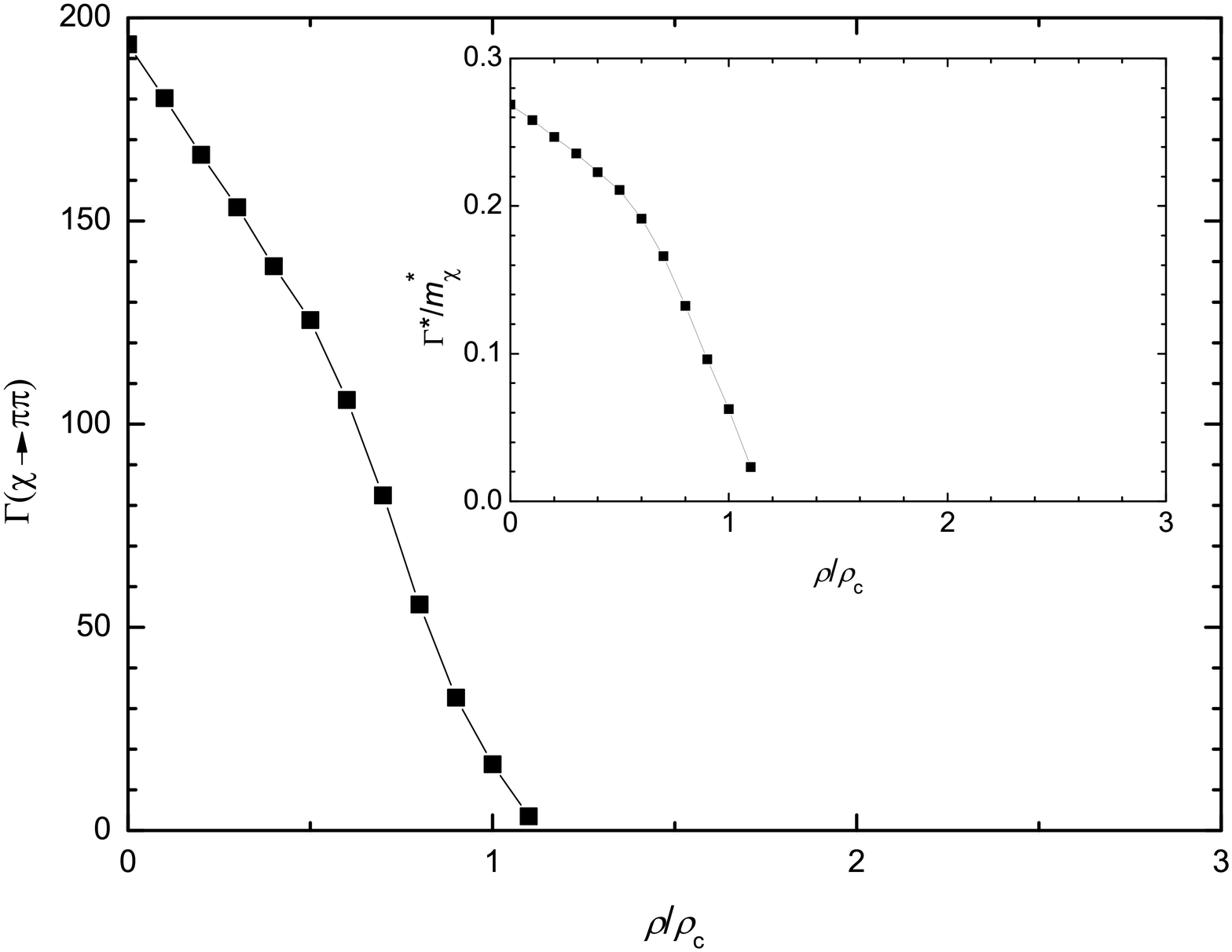,width=6cm,angle=0}}
\centerline{(a) \hskip 6cm (b)}
\caption{(a) The ratios of the in-medium parameters to the free space
parameters. The graph in a small box shows the masses of the pion
and the scalar, (b) the in-medium decay width $\Gamma^*(\chi
\rightarrow\pi\pi)$ as a function of $\rho$.}
\label{pion-in-medium}
\end{figure}

In Fig. \ref{pion-in-medium} we show the (exact) ratios of the in-medium parameters
relative to their free-space values. Only the results obtained with the parameter
set B are shown. The parameter set A yields similar results while set C
shows a two step structure with an intermediate pseudogap phase.
Not only the average value of $\chi_0$ over the space but also $\chi_0(\vec{r})$ itself
vanishes at any point in space.
This is the reason for the vanishing of $m_\pi^*$ and $f_\pi^*$.
That is, $f_\pi^*$ really vanishes when $\rho < \rho_c$  in the Skyrme model with dilaton dynamics.

At low matter density,  the ratio $f_\pi^*/f_\pi$ can be
fitted to a linear function
\begin{equation}
\frac{f_\pi^*}{f_\pi} \sim 1 - 0.24(\rho/\rho_0) + \cdots
\end{equation}
At $\rho=\rho_0$, this yields $f_\pi^*/f_\pi=0.76$, which is comparable to the other
predictions.

In Ref. \refcite{Lee:2003rj}, the in medium modification of the $\chi$ decay into two pions
is also studied using  the mean field approximation.
Gathering the terms with a fluctuating scalar field and two fluctuating pion
fields, we get the Lagrangian density for the process
$\chi \rightarrow \pi\pi$
\begin{equation}
{\cal L}_{M,\chi\pi^2} = \frac{\chi_0}{f_\chi^2}
(\delta_{ab}+g_{ab}) \chi \partial_\mu \phi_a \partial^\mu \phi_b,
\end{equation}
where only the term from ${\cal L}_\sigma$ is used.
Averaging the space dependence of the background field
configuration modifies the coupling constant by a factor $\langle
(\chi_0/f_\chi)(1+g_{11})\rangle = \langle
(\chi_0/f_\chi)(1-\frac23\pi^2)\rangle $. Taking into account the
appropriate wave function renormalization factors, $Z_\pi$, and
the change in the scalar mass, we obtain the in-medium
decay width  as
\begin{equation}
\Gamma^* (\chi \rightarrow \pi\pi) =
 \frac{3m_\chi^{*3}}{32\pi f_\chi^2}
 \left| \frac{\langle(\chi_0/f_\chi)(1-\frac23\pi^2)\rangle}
{\langle(\chi_0/f_\chi)^2 (1-\frac23\pi^2)\rangle} \right|^2
\approx \frac{3 m_\chi^{*3}}{32\pi f_\chi^{*2}}.
\label{exact}\end{equation}

We show in Fig. \ref{pion-in-medium} the in-medium decay width
predicted with the parameter set B. In the region $\rho \geq
\rho_{pt}$ where $\chi_0=0$, $\Gamma^*$ cannot be defined to this
order. Near the critical point, the scalar becomes an extremely
narrow-width excitation, a feature which has been discussed in the
literature as a signal for chiral restoration.~\cite{Hatsuda:2001da,Fujii:2003bz}

Another interesting change in the properties of the pion in the medium is
associated with the in medium pion dispersion relation. This relation requires, besides the mass,
the so-called in medium pion
velocity, $v_\pi$. This property allows us to gain more insight
into the real time properties of the system under extreme
conditions and enables us to analyze how the phase transition from
normal matter to deconfined QCD takes place
from the hadronic side , the so called `bottom up' approach.

At nonzero temperature and/or density,  the Lorentz symmetry is broken by the medium. In the
dispersion relation for the pion modes (in the chiral limit)
 \begin{equation}
p_0^2 = v_\pi^2 |\vec{p}|^2,
 \label{disp_rel}
\end{equation}
the velocity $v_\pi$ which is 1 in free-space must depart from 1.
This may be studied reliably, at least at low temperatures and at
low densities, via chiral perturbation theory.~\cite{Pisarski:1996mt} The
in-medium pion velocity can be expressed in terms of the time
component of the pion decay constant, $f^t_\pi$ and the space
component, $f_\pi^s$,~\cite{Leutwyler:1993gf,Kirchbach:1997rk}
\begin{equation}
\begin{array}{l}
\langle 0 | A^0_a | \pi^b(p) \rangle_{\mbox{\scriptsize in-medium}}
= i f^t_\pi \delta^{ab} p^0, \\
\langle 0 | A^i_a | \pi^b(p) \rangle_{\mbox{\scriptsize
in-medium}} = i f^s_\pi \delta^{ab} p^i.
\end{array}
\label{decay_consts}
 \end{equation}
The conservation of the axial
vector current leads to the dispersion relation (\ref{disp_rel})
with the pion velocity given by
 \begin{equation} v_\pi^2 = f_\pi^s /f_\pi^t.\label{vpi1}
  \end{equation}

In Ref. \refcite{Son:2002ci} two decay constants, $f_t$ and
$f_s$, are defined differently from those of Eq.(\ref{decay_consts})), through
the effective Lagrangian,
\begin{equation}
{\cal L}_{\mbox{\scriptsize eff}} = \frac{f_t^2}{4} \mbox{Tr} (\partial_0
U^{\dagger} \partial_0 U ) - \frac{f_s^2}{4} \mbox{Tr} (\partial_i U^{\dagger}
\partial_i U ) + \cdots, \label{Leff}
 \end{equation}
where $U$ is an SU(2)-valued chiral field whose phase describes
the in-medium pion. In terms of these constants, the pion velocity
is defined by
 \begin{equation}
 v_\pi=f_s/f_t.\label{vpi2}
 \end{equation}

\begin{figure}
\centerline{\epsfig{file=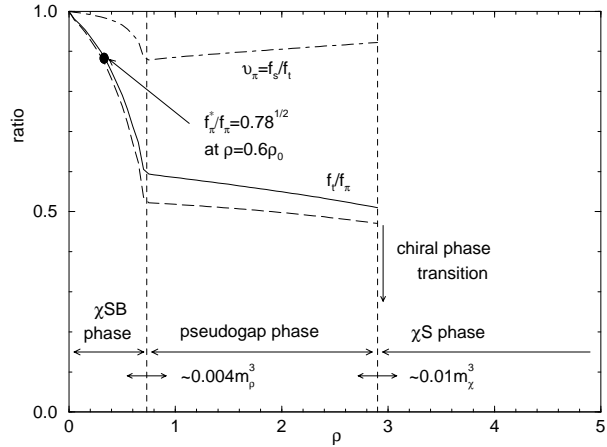,height=8cm,angle=270}} \
\caption{In-medium pion decay constants and their ratio, the pion
velocity.}
\label{vpi}
\end{figure}

In Ref. \refcite{Lee:2003rj}, it is shown that local interactions with background
skyrmion matter lead to a breakdown of Lorentz symmetry in the dense medium and to an effective
Lagrangian for pion dynamics in the form of Eq.(\ref{Leff}).
The results are shown in Figs.~\ref{vpi}. Both of the
pion decay constants change significantly as a function of density
and vanish -- in the chiral limit -- when chiral symmetry is
restored. However, the second-order contributions to the $f_s$ and
$f_\pi$, which break Lorentz symmetry, turn out to be rather
small, and thus their ratio, the pion velocity, stays $v_\pi \sim
1$. The lowest value found is $\sim 0.9$. Note, however, the
drastic change in its behavior at two different
densities. At the lower density, where  skyrmion matter is in
the chiral symmetry broken phase, the pion velocity decreases and
has the minimum at $\rho = \rho_p$. If one worked only at low
density in a perturbative scheme, one would conclude that the pion
velocity decreases all the way to zero. However, the presence of
the pseudogap phase transition changes this behavior. In the
pseudogap phase, the pion velocity not only stops decreasing but
starts increasing with increasing density. In the chiral
symmetry restored phase both $f_t$ and $f_s$ vanish.
Thus  their ratio makes no sense.

In Refs.~\refcite{Kalloniatis:2004fx,Kalloniatis:2005kc}, the in-medium
modification of the neutral pion decay process into two gammas and
neutrino-anti-neutrino pair are studied in the same manner.
The $\pi^0 \rightarrow \gamma \gamma$ process is shown to be strongly
suppressed in dense medium, while the process $\pi^0 \rightarrow \nu\bar{\nu}$
forbidden in free space becomes possible by the Lorentz symmetry breaking
effect of the medium.

\section{Skyrmion matter at finite temperature }\label{hotSK}

There are many studies of  lattice QCD at finite temperature.
The situation is completely different for skyrmion matter where
the number of studies is limited. For example, skyrmion matter
has been heated up to melt the crystal into a
liquid to study the crystal-liquid phase transition.~\cite{Kaelbermann:1998wp,Schwindt:2002we}
However this  phenomenon
 is irrelevant to study the restoration of chiral symmetry, which interests  us
 here for the reasons discussed in previos sections.

What happens if we  heat up the system?  Naively, as the temperature increases,
the kinetic energy of the skyrmions increases and the skyrmion crystal
begins to melt. The kinetic energy associated with the translations, vibrations and
rotations of the skyrmions is proportional to $T$. This mechanism
leads to a solid-liquid-gas phase transition of the skyrmion system.
However, we are interested in the chiral symmetry restoration transition,
which is not related to the melting.
Therefore,  a new mechanism must be incorporated
to describe chiral symmetry restoration. We  show in what follows that the thermal excitation of the pions
 in the medium is the appropriate mechanism, since
this phenomenon is proportional to $T^4$ and therefore
dominates the absorption of heat.

The pressure of non-interacting pions is given by ~\cite{Bochkarev:1995gi}
\begin{equation}
P=\frac{\pi^2}{30} T^4,
\end{equation}
where we have taken into account the contributions from three species of pion,
$\pi^+, \pi^0, \pi^-$.
This term contributes to the energy per single skyrmion volume as $3PV (\chi/f_\chi)^2$.
The kinetic energy of the pions arises from ${\cal L}_\sigma$ (\ref{kinetic}),
and therefore scale symmetry
implies that it should carry a factor $\chi^2$.
The factor 3 comes from the fact that our pions are massless.

To estimate the properties of skyrmion matter at finite temperature let us take $\chi$ as a constant field
as we did in Sec. \ref{denseSK_dilaton}.  After including
thermal pions, Eq. (\ref{E_over_B}) can be rewritten as
\begin{equation}
E/B(\rho, T, X) =
\left(E_2/B)(\rho) + \frac{\pi^2}{10}T^4 V \right) X^2
+ (E_4/B)(\rho) + X^4 (\ln X - \textstyle \frac14)+\frac14 ),
\label{E1}\end{equation}
where we have dropped the pion mass term.

As in Sec. \ref{denseSK_dilaton}, chiral restoration will occur when the value of
$X_{\rm min}$ that minimizes $E/B$ vanishes.
By minimizing $E/B$ with respect to $X$, we observe that
the phase transition  from a non-vanishing
$X= e^{-1/4}$ to $X= 0$.
Thus, the nature of the phase transition is of the first order.

After a straightforward calculation we obtain,
\begin{equation}
\rho^c (E_2/B)  + \frac{\pi^2}{10} T_c^4  = \frac{f_\chi^2
m_\chi^2}{8e^{1/2}} .
\label{PTRT}\end{equation}
which leads to
\begin{equation}
T_c =  \left( \frac{10}{\pi^2} \left(\frac{f_\chi^2
m_\chi^2}{8e^{1/2}}
 -\rho^c(E_2/B)(\rho^c)\right) \right)^{1/4}
\label{Tc}\end{equation}
For $\rho=0$ (zero density),
our estimate for the critical temperature is
\begin{equation}
T_c =  \left( \frac{10}{\pi^2} \frac{f_\chi^2 m_\chi^2}{8e^{1/2}}
\right)^{1/4} \sim  \mbox{205 MeV},
\end{equation}
where we have used  the following values for the parameters. $f_\chi=210$ MeV and $m_\chi=720$ MeV.
It is remarkable that our  model leads to $T_c
\sim 200$ MeV, which is close to that obtained by lattice QCD ~\cite{Karsch:2007dt}
and in agreement with the data.~\cite{Arsene20051} To us this is a confirmation that the mechanism
chosen for the absorption of heat plays a fundamental role in the hadronic phase.

\begin{figure}[tbp]
\centerline{\epsfig{file=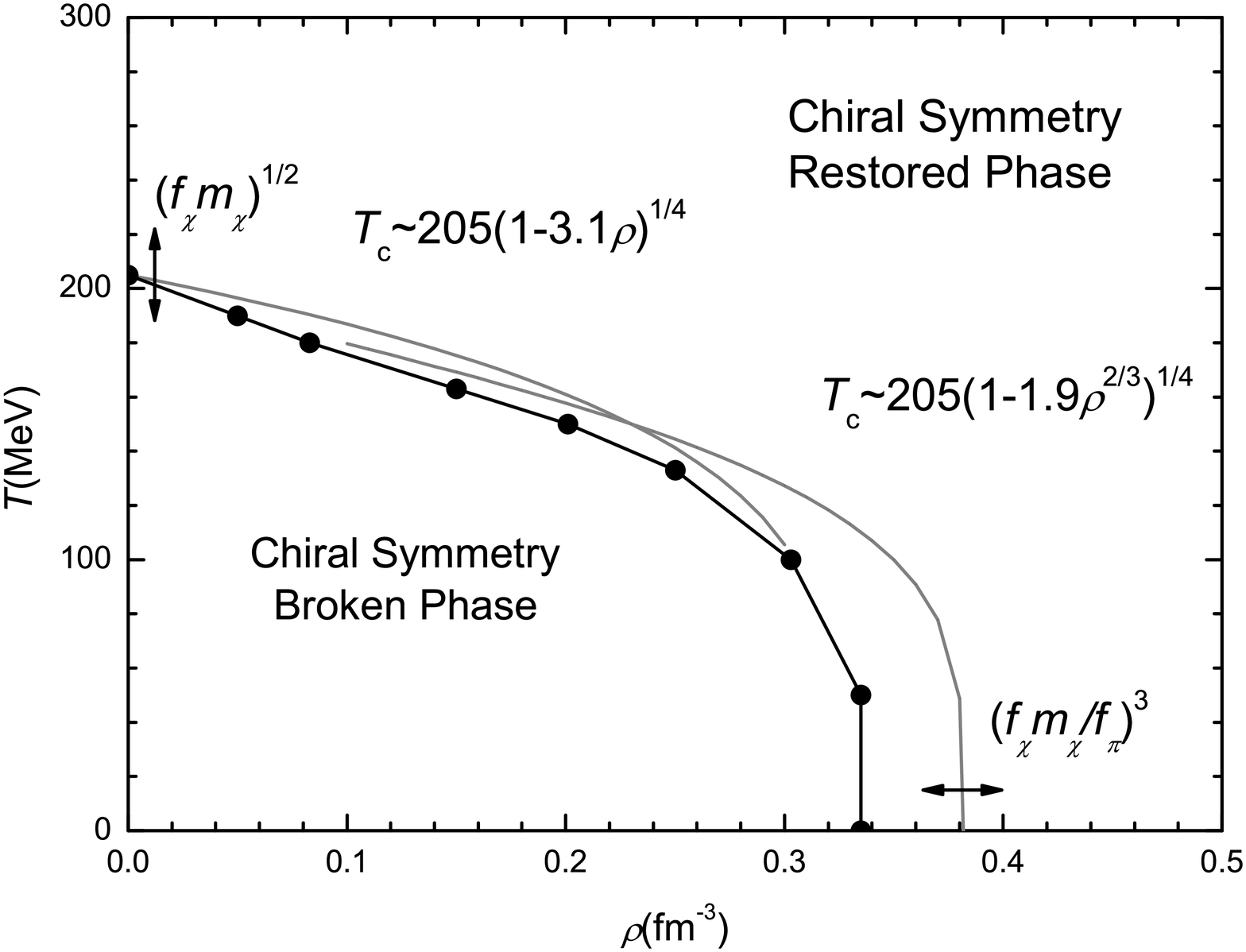,width=8cm,angle=0}}
\caption{The chiral phase transition. The solid line shows the exact calculation,
while the gray lines two approximate estimates. }
\label{phasediagramT}
\end{figure}

The numerical results on $E_2/B$ that minimize the energy of the system for a given $\rho_B$
can be approximated by
\begin{equation}
E_2/B = \left\{
\begin{array}{ll}
\displaystyle 10 f_\pi^2 /\rho^{1/3},  &
\rho > \rho_0 \\
\displaystyle 36 f_\pi / e_{\mbox{\scriptsize sk}} , &
\rho< \rho_0,
\end{array}
\right. \label{Asy}\end{equation} where
$\rho_0 =(e_{\mbox{\scriptsize sk}}f_\pi/3.6)^3$.

Using Eq.(\ref{Asy}) for $E_2/B$, we obtain the critical density for chiral
symmetry restoration at zero temperature as
\begin{equation}
\rho^c (T=0) = \left( \frac{f_\chi^2 m_\chi^2 }{8e^{1/2}}
\frac{1}{10f_\pi^2} \right)^{3/2} \sim 0.37 \mbox{ fm}^{-3}.
\label{Rho}
\end{equation}
Since $\rho_0 = 0.24$ fm$^{-3} < \rho^c(T=0) $ our result is consistent with
the high density formula for $E_2/B$ used.

The resulting critical density $\rho^c(T=0) \sim 0.37$ fm$^{-3}$ is
only twice  normal nuclear matter density and it is
low with respect to the expected values.
This result does not represent a problem since  $\rho^c (T=0)$ scales with $(f_\chi m_\chi / f_\pi)^3$
and $T_c^{\rho=0}$ with $(f_\chi m_\chi )^{1/2}$ and small changes in the parameters  lead to
larger values  for the critical density without changing the critical temperature too much.

For a finite density smaller than $\rho^c (T=0)$, we obtain the
corresponding critical temperature by substituting the asymptotic
formulas (\ref{Asy}) for $E_2/B$,
\begin{equation}
T_c = \left\{
\begin{array}{lcl}
\displaystyle T_c(\rho=0)\;(1 - 3.09\; \rho_c)^{1/4} & \mbox{for} & \rho < \rho_0, \\
\displaystyle T_c(\rho=0)\;(1 - 1.92\; \rho_c^{2/3})^{1/4}  & \mbox{for} & \rho > \rho_0,
\end{array}
\right.
\end{equation}
where the density is measured in fm$^{-3}$
The gray lines in Fig. \ref{phasediagramT} show these two curves.
The results from the {\em exact} calculations obtaned by minimization of the energy (\ref{E1})
are shown by black dots connected by black line
in Fig. \ref{phasediagramT}.
The resulting phase diagram  has the same shape but the values of the temperatures
and densities are generally smaller than in the approximate estimates.

\section{Vector mesons and dense  matter} \label{vectorSK}
In our effort to approach the theory of the hadronic interactions and inspired
by Weinberg's theorem ~\cite{Weinberg:1978kz} we proceed to incorporate
to the model the lowest-lying vector mesons, namely the $\rho$ and the $\omega$.
In this way we also  do away with the ad hoc
Skyrme quartic term. It
is known that these vector mesons play a crucial role in
stabilizing the single nucleon system ~\cite{Zahed:1986qz,Meissner:1987ge} as well as
in the saturation of normal nuclear matter.~\cite{Serot:1984ey}

We consider a skyrmion-type Lagrangian with vector mesons
possessing hidden local gauge symmetry,
~\cite{Bando:1987br} spontaneously broken chiral
symmetry and scale symmetry.~\cite{Ellis:1984jv,Lee:2003eg}
Such a theory might be considered as
a better approximation to reality than the extreme large $N_c$
approximation to QCD represented by the Skyrme model.
Specifically, the model Lagrangian, which we investigate, is given by ~\cite{Meissner:1999pe}

\begin{eqnarray}
{\cal L} &=& \frac{f_\pi^2}{4} \left(\frac{\chi}{f_\chi}\right)^2
\mbox{Tr}(\partial_\mu U^\dagger \partial^\mu U) + \frac{f_\pi^2
m_\pi^2}{4} \left(\frac{\chi}{f_\chi}\right)^3
    \mbox{Tr}(U+U^\dagger-2)
\nonumber\\
&&
-\frac{f_\pi^2}{4} a \left(\frac{\chi}{f_\chi}\right)^2
 \mbox{Tr}[\ell_\mu + r_\mu + i(g/2)
( \vec{\tau}\cdot\vec{\rho}_\mu + \omega_\mu)]^2
-\textstyle \frac{1}{4} \displaystyle
\vec{\rho}_{\mu\nu} \cdot \vec{\rho}^{\mu\nu}
-\textstyle \frac{1}{4}  \omega_{\mu\nu} \omega^{\mu\nu}
\nonumber\\
&& +\textstyle\frac{3}{2} g \omega_\mu B^\mu
+\textstyle\frac{1}{2} \partial_\mu \chi \partial^\mu \chi
-\displaystyle \frac{m_\chi^2 f_\chi^2}{4} \left[ (\chi/f_\chi)^4
(\mbox{ln}(\chi/f_\chi)-\textstyle\frac14) + \frac14 \right],
\label{lag}\end{eqnarray}
where, $U =\exp(i\vec{\tau}\cdot\vec{\pi}/f_\pi) \equiv \xi^2$,
$\ell_\mu = \xi^\dagger \partial_\mu \xi$,
$r_\mu = \xi \partial_\mu \xi^\dagger$, $
\vec{\rho}_{\mu\nu} = \partial_\mu \vec{\rho}_\nu
- \partial_\nu \vec{\rho}_\mu + g \vec{\rho}_\mu \times \vec{\rho}_\nu$,
$\omega_{\mu\nu}=\partial_\mu\omega_\nu-\partial_\nu\omega_\mu$, and
$B^\mu =  \frac{1}{24\pi^2} \varepsilon^{\mu\nu\alpha\beta}
\mbox{Tr}(U^\dagger\partial_\nu U U^\dagger\partial_\alpha U
U^\dagger\partial_\beta U)$.
Note that the Skyrme quartic term is not present. The
vector mesons, $\rho$ and $\omega$, are incorporated as dynamical
gauge bosons for the local hidden gauge symmetry of the non-linear
sigma model Lagrangian and the dilaton field $\chi$ is introduced
so that the Lagrangian has the same scaling behavior as QCD. The
physical parameters appearing in the Lagrangian are summarized in
Table \ref{parametersVM} .

\begin{table}[h]
\tbl{Parameters of the model Lagrangian}
{\begin{tabular}{ccc}
\toprule
notation & physical meaning & value \\
\colrule
$f_\pi$ & pion decay constant & 93 MeV \\
$f_\chi$ & $\chi$ decay constant & 210 MeV \\
$g$ & $\rho\pi\pi$ coupling constant &
5.85$^*$ \\
$m_\pi$  & pion mass & 140 MeV \\
$m_\chi$ & $\chi$ meass & 720 MeV \\
$m_V$ & vector meson masses &
770 MeV$^\dagger$  \\
$a$ & vector meson dominance & 2 \\
\botrule
\multicolumn{3}{l}{\tablefont $^*$ obtained
by using the KSFR relation $m_V^2=m_\rho^2=m_\omega^2=af_\pi^2
g^2$ with }\\
\multicolumn{3}{l}{\tablefont \hskip 1em $a=2$.
cf. $g_{\rho\pi\pi}=6.11$ from the decay width of
$\rho\rightarrow\pi\pi$.} \\
\multicolumn{3}{l}{\tablefont $^\dagger$ experimentally measured values
are $m_\rho$=768 MeV and $m_\omega$=782 MeV.}
\end{tabular}}
\label{parametersVM}
\end{table}

\subsection{Dynamics of the single skyrmion}

The spherically symmetric hedgehog Ansatz for the
$B=1$ soliton solution of the standard Skyrme model can be
generalized to
\begin{equation}
U^{B=1} = \exp(i\vec{\tau}\cdot\hat{r} F(r)),
\label{UBeq1}\end{equation}
\begin{equation}
\rho^{a,B=1}_{\mu=i} = \varepsilon^{ika}\hat{r}^k \frac{G(r)}{gr},\hskip 2em
\rho^{a,B=1}_{\mu=0} = 0,
\label{RBeq1}\end{equation}
\begin{equation}
\omega^{B=1}_{\mu=i} = 0,\hskip 2em
\omega^{B=1}_{\mu=0} = f_\pi W(r),
\label{WBeq1}\end{equation}
\begin{equation}
\chi^{B=1} = f_\chi C(r).
 \label{CBeq1}\end{equation}
The boundary conditions that the profile functions satisfy at
infinity are
 \begin{equation}
F(\infty)=G(\infty)=W(\infty)=0, \hskip 2em
C(\infty)=1,
\label{BC_infty}
\end{equation}
and at the center ($r=0$) are
\begin{equation}
F(0)=\pi, \hskip 2em G(0) = -2, \hskip 2em W^\prime(0)=C^\prime(0)=0.
\label{BC_0}
\end{equation}

\begin{figure}[tbp]
\centerline{\epsfig{file=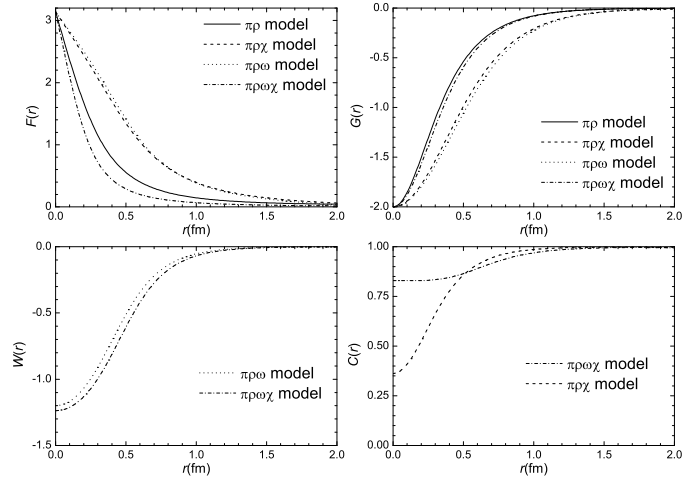,width=9cm,angle=0}}
\caption{Profile functions - $F(r)$, $G(r)$, $W(r)$ and $C(r)$.}
\label{profilesVM}
\end{figure}

The profile functions are
obtained numerically by minimizing the soliton mass with the boundary conditions
(see Ref. \refcite{Park:2003sd} for the technical details).
The results are summarized in Table \ref{massVM} and the corresponding
profile functions are given in Fig. \ref{profilesVM}.
The role of the $\omega$ meson that provides a strong repulsion is prominent.
Comparing the $\pi \rho$ model with the
$\pi \rho \omega$ model, the presence of the
$\omega$ increases the mass by more than 415 MeV and the size,
i.e. $\langle r^2\rangle$, by more than .28 fm$^2$.

How does the dilaton affect this calculation? The $\pi \rho$ model
with much smaller skyrmion has a larger baryon density near the
origin and this affects the dilaton, significantly changing its
mean-field value from its vacuum one.  The net effect of the
dilaton mean field on the mass is a reduction of $\sim$ 150 MeV,
whereas for the $\pi \rho \omega$ model it is only of 50 MeV. The
details can be seen in Table \ref{massVM}. The effect on the soliton size is,
however, different: while the dilaton in the $\pi \rho$ model
produces an additional localization of the baryon charge and hence
reduces $\langle r^2\rangle$ from .21 fm$^2$ to .19 fm$^2$, in the $\pi
\rho \omega$ model, on the contrary, the dilaton produces a
$delocalization$ and increases $\langle r^2\rangle$ from .49 fm$^2$ to .51
fm$^2$. We will see, however, that this strong repulsion provided by $\omega$ causes a
somewhat serious problem in the chiral restoration of the skyrmion matter
at higher density.

\begin{table}
\tbl{Single skyrmion mass and various contributions to it.}
{\begin{tabular}{ccccccccc}
\toprule
Model & $\langle r^2\rangle$ & $E^{B=1}$ & $E^{B=1}_\pi$ & $E^{B=1}_{\pi\rho}$
& $E^{B=1}_\rho$ & $E^{B=1}_\omega$ & $E^{B=1}_{WZ}$ & $E^{B=1}_\chi$ \\
\colrule
$\pi\rho$-model & 0.27 & 1054.6 & 400.2 + 9.2 & 110.4 & 534.9
& 0.0 & 0.0 & 0.0 \\
$\pi\rho\chi$-model & 0.19 & 906.5 & 103.1 + 1.4 & 155.1 & 504.1
& 0.0 & 0.0 & 142.8 \\
$\pi\rho\omega$-model & 0.49 & 1469.0 & 767.6 + 39.9 & 33.2 & 370.7
  & -257.6 & 515.1 & 0.0 \\
$\pi\rho\omega\chi$-model & 0.51 & 1408.3 & 646.0 + 29.2 & 34.9 & 355.7
  & -278.3 & 556.7 & 64.2 \\
\botrule
\end{tabular}}
\label{massVM}
\end{table}

\subsection{Skyrmion Matter : an FCC skyrmion crystal}

Again, the lowest-energy configuration is obtained when one of the
skyrmions is rotated in isospin space with respect to the other by
an angle $\pi$ about an axis perpendicular to the line joining the
two.~\cite{Park:2003sd} If we generalize this Ansatz to many-skyrmion matter, we
obtain that the configuration at the classical level for a given
baryon number density is an FCC crystal where the nearest neighbour
skyrmions are arranged to have the attractive relative
orientations.~\cite{Lee:2003aq}
Kugler's Fourier series expansion method~\cite{Kugler:1989uc} can be generalized
to incorporate the vector mesons, although some subtelties associated with the
vector fields have to be implemented.
The details can be found in Ref. \refcite{Park:2003sd}.

\begin{figure}[tbp]
\centerline{\epsfig{file=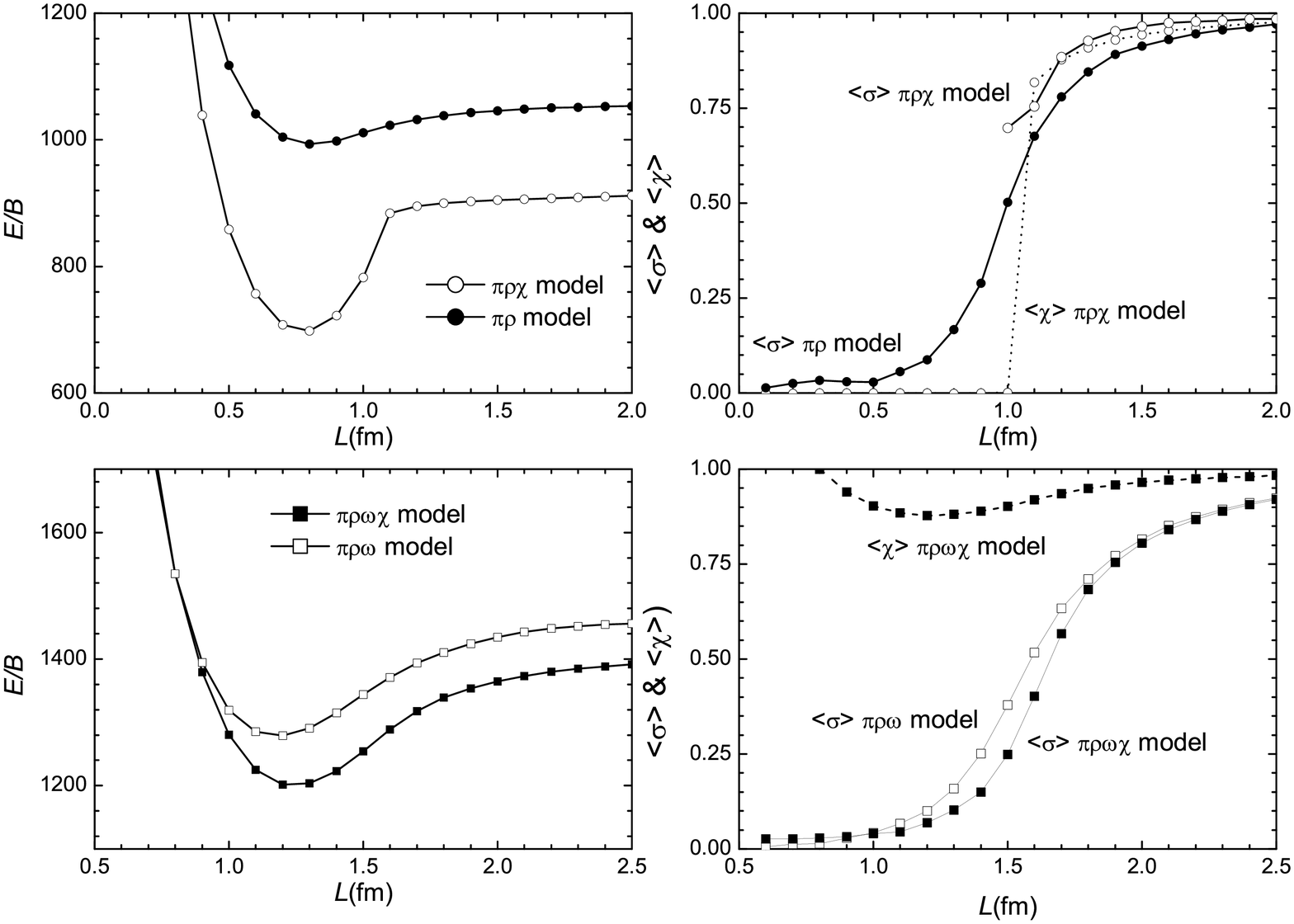,width=12cm,angle=0}}
\caption{$E/B$,
$\langle \chi\rangle$ and $\langle\sigma\rangle$ as a function of
$L$ in the models (a) without the $\omega$ and (b) with omega.}
\label{VM_profile}
\end{figure}

Figs.~ \ref{VM_profile} are the the numerical results of the energy
per baryon $E/B$, $\langle \chi\rangle$ and $\langle
\sigma\rangle$ in various models as a function of the FCC lattice
parameter $L$.
In the $\pi \rho \chi$ model, as the density of the system
increases ($L$ decreases), $E/B$ changes little. It
is close to the energy of a $B=1$ skyrmion up to a density greater
than $\rho_0 \; (L \sim 1.43)$. This result is easy to interpret.
As we discussed before the size of the skyrmion in this model
is very small and therefore the skyrmions in the lattice will
interact only at very high densities, high enough for their tails
to overlap.

In the absence of the $\omega$, the dilaton field plays a dramatic
role. A skyrmion matter undergoes an abrupt phase transition at
high density at which the expectation value of the dilaton field
vanishes $\langle \chi \rangle=0$. (In general, $\langle
\chi\rangle =0$ does not necessarily require $\langle \chi^2
\rangle =0$. However, since $\chi \geq 0$, $\langle
\chi\rangle =0$ always accompanies $\chi=0$ in the whole space.)
The $\rho$ meson on the other hand is basically a
spectator at the classical level, producing little change with
respect to our previously studied $\pi \chi$ model
 except that at high densities, once the
$\rho$ starts to overlap, the energy of nuclear matter increases
due to its the repulsive effect at short distances. The densities
have to be quite high since these skyrmions are very small. Since
$\chi$ vanishes at the phase transition, we recover the standard behavior,
namely, $f_\pi^*=0$  and $m_\rho^*=0$.

In the  $\pi \rho \omega \chi$ model, the situation changes dramatically.
The reason is that the
$\omega$ provides not only a strong repulsion among the skyrmions,
but somewhat surprisingly, also an intermediate range attraction.
Note the different mass scales between Figs.~ \ref{VM_profile}(a)
and \ref{VM_profile}(b). In both the
$\pi\rho\omega$ and the $\pi\rho\omega\chi$ models, at high
density, the interaction reduces $E/B$ to 85\% of the $B=1$
skyrmion mass. This value should be compared with 94\% in the
$\pi\rho$ model. In the $\pi\rho\chi$-model, $E/B$ goes down to
74\% of the $B=1$ skyrmion mass, but in this case it is due to the
dramatic behavior of the dilaton field.

In the $\pi\rho\omega\chi$ model the role of the dilaton field is
suppressed. It provides a only a small attraction at intermediate
densities. Moreover, the phase transition towards its vanishing
expectation value, $\langle \chi\rangle=0$, does not take place.
Instead, its value grows at high density!

The problem involved is associated with the Lagrangian (\ref{lag})
which includes an anomalous part  known as Wess-Zumino term, namely
 the coupling of the $\omega$ to the Baryon current, $B_\mu$.
To see that this term is the one causing the problem, consider the energy per baryon
contributed by this term.~\cite{Park:2003sd}
 \begin{equation}
 \left(\frac{E}{B}\right)_{WZ}=\frac 14(\frac{3g}{2})^2\int_{Box}
 d^3x\int d^3x^\prime B_0 (\vec{x})
 \frac{{\rm exp}(-m_\omega^*|\vec{x}-\vec{x}^\prime|)}
 {4\pi|\vec{x}-\vec{x}^\prime|} B_0 (\vec{x}^\prime)\label{wzenergy}
 \end{equation}
where ``Box" corresponds to a single FCC cell.
Note that while the integral over $\vec{x}$
is defined in a single FCC cell, that over $\vec{x}^\prime$ is
not. Thus, unless it is screened, the periodic source $B_0$
filling infinite space will produce an infinite potential $w$
which leads to an infinite $(E/B)_{WZ}$. The screening is done by
the omega mass, $m_\omega^*$. Thus the effective $\omega$ mass cannot vanish.
Our numerical results reflect this fact: at high
density the $B_0$-$B_0$ interaction becomes large compared to any
other contribution. In order to reduce it, $\chi$ has to increase,
and thereby the effective screening mass $m_\omega^* \sim m_\omega
\langle\chi\rangle$ becomes larger. In this way we run into a
phase transition where the expectation value of $\chi$ does not
vanish and therefore $f_\pi$ does not vanish but instead
increases.

\subsection{A Resolution of the $\omega$ problem}

Assuming that there is nothing wrong with (\ref{lag}), we focus on
the Wess-Zumino term in the Lagrangian. Our objective is to find an
alternative to (\ref{lag}) that leads to a behavior consistent
with the expected behavior. In the absence of any
reliable clue, we try the simplest, admittedly {\it ad hoc},
modification of the Lagrangian (\ref{lag}) that allows a
reasonable and appealing way-out.~\cite{Park:2008zg}  Given our ignorance as to how
spontaneously broken scale invariance manifests in matter,  we shall
simply forego the requirement that the anomalous term be scale
invariant and multiply the anomalous $\omega\cdot B$ term by $(\chi/f_\chi)^n$
for $n\ge 2$. We have verified that it matters little whether we
pick $n=2$ or $n=3$.~\cite{Park:2008zg} We therefore take $n=3$:

\begin{equation}
L_{an}^\prime = \textstyle\frac{3}{2} g (\chi/f_\chi)^3 \omega_\mu
B^\mu \label{lag-anp}\end{equation}
This additional factor has two virtues:

\begin{itemize}
\item [ i)] It leaves meson dynamics in free space
(i.e. $\chi/f_\chi = 1$) unaffected,
since chiral symmetry is realized \`a la sigma model as
required by QCD.

\item [ii)] It plays the role of an effective density-dependent
coupling constant so that at high density, when scale symmetry is
restored and $\chi/f_\chi \rightarrow 0$,  there will be no coupling
between the $\omega$ and the baryon density as required by hidden local
symmetry with the vector manfiestation.

\end{itemize}

The properties of this Lagrangian for the meson ($B=0$) sector are
the same as in our old description. The parameters of the Lagrangian
are determined by meson physics as given in Table \ref{parametersVM}

\begin{figure}[tbp]
\centerline{\epsfig{file=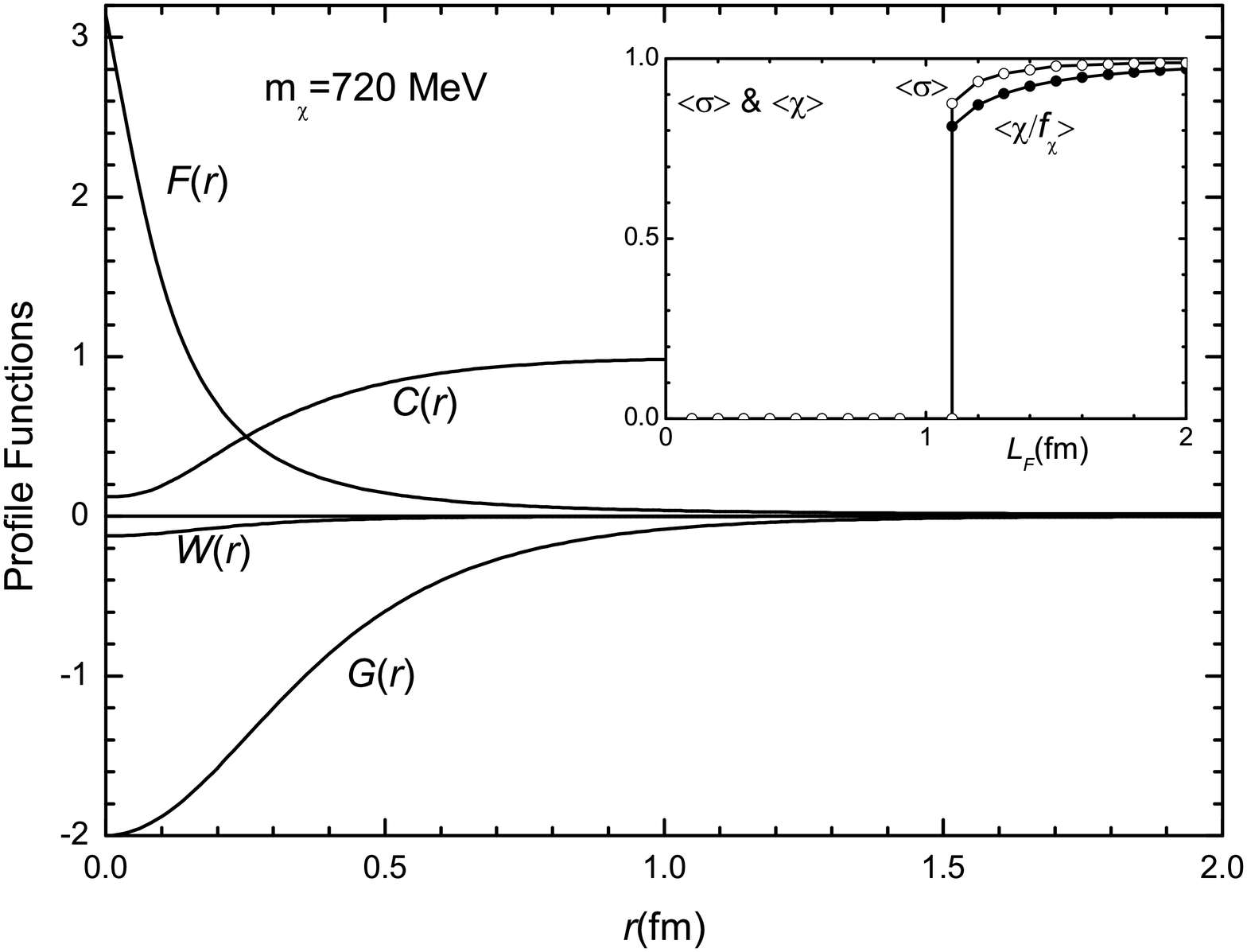,width=6cm}\epsfig{file=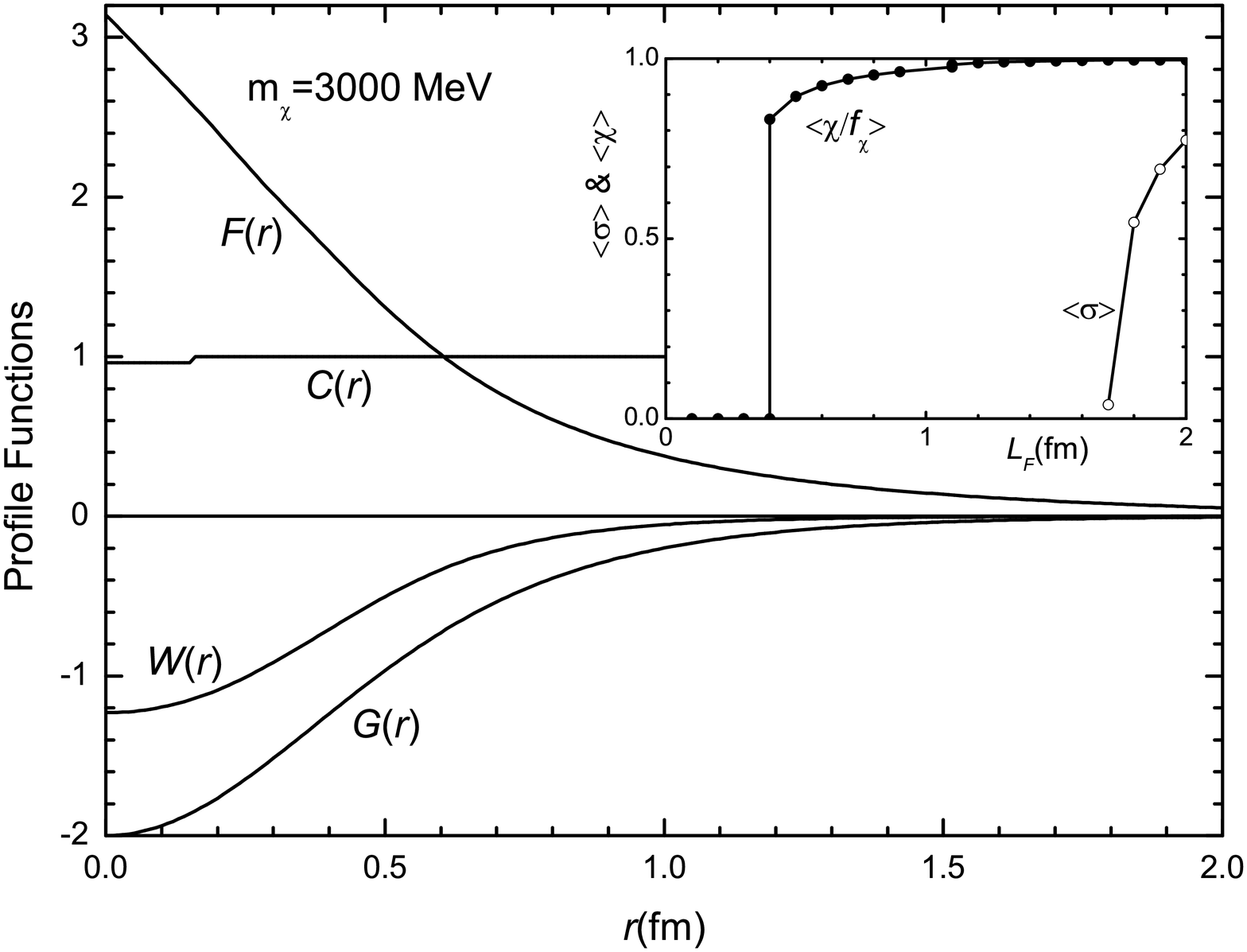,width=6cm}}
\caption{A small and large skyrmion obtained with $m_\chi=720$MeV (left) and $m_\chi=3000$ MeV.
Shown in small boxes are $\langle \chi\rangle$ and $\langle \sigma\rangle$ as a function of the FCC lattice size
$L_F$. \protect\label{VM_resolved}}
\end{figure}

Fig. \ref{VM_resolved} summarizes the consequences of the modification.
Depending on the dilaton mass, the properties of a single skyrmion show distinguished characters
and consequently undergoes different phase transition.
A small dilaton mass, say $m_\chi \sim 1$ GeV, leads to a very small skyrmion with an rms radius
about $0.1$ fm. The light dilaton seems to react quite sensitively to the
presence of the matter. One can see that at the center of the single skyrmion the chiral symmetry
is almost restored. It weakens most of the repulsion from $\omega-B$ coupling, which leads us
to such a small sized skyrmion. Since these small skyrmions are already chiral-symmetry-restored objects,
simply filling the space with them restores the symmetry. As shown in the small box,
chiral symmetry is restored simultaneously when $\langle \sigma \rangle$ vanishes.
In case of having a large mass, the dilaton does not play any significant role in the structure of a
 single skyrmion.  This scenario leads, as  the density of skyrmion matter increases, first to a pseudogap phase transition
where $\langle \sigma \rangle=0$ and thereafter, at higher density, to a genuine chiral symmetry restoration phase transition
where $\langle \chi/f_\chi \rangle=0$.
Anyway, whether the dilaton is light or heavy, we finally have a reasonable phase transition scenario
that at some critical density  chiral symmetry restoration occurs where $\langle \chi/f_\chi\rangle$ vanishes.

 Under the same mean field approximation, this skyrmion approach to the dense matter leads us to
the scaling behaviors of the vector mesons
\begin{equation}
\frac{m_\rho^*}{m_\rho} = \frac{m_\omega^*}{m_\omega}
= \sqrt{ \langle \left(\frac{\chi}{f_\chi} \right)^2 \rangle},
\end{equation}
while that of the pion decay constant is
\begin{equation}
\frac{f_\pi^*}{f_\pi} = \sqrt{ \langle \left(\frac{\chi}{f_\chi} \right)^2 \textstyle
(1+(a-1)\frac23 \pi^2) \displaystyle \rangle}.
\end{equation}
With $a=1$, a remarkably simple BR scaling law is obtained.
These scaling laws imply that as the density of the matter increases the effecive
quantities in medium scale down.

We have shown how a slight modification of the Lagrangian resolves the $\omega$ problem.
However, in modifying the Lagrangian we have taken into account only the phenomenological side
of the problem. Multiplying the Wess-Zumino term by the factor $(\chi/f_\chi)^n$ has
no sound theoretical support. It breaks explicitly the scale invariance of the Lagrangian.
Recall that the dilaton field was introduced into the model to respect  scale symmetry.
Furthermore, we don't have any special reason for choosing $n=3$, except that it works well.
Recently, a more fundamental explanation for the behavior in Eq.(\ref{lag-anp}) has been found.
\footnote{Private communication by M. Rho on work in progress by H.K. Lee and M. Rho.}

\section{Conclusions} \label{concl}

In trying to understand what happens to hadrons under extreme conditions,
it is necessary that the theory adopted for the description be consistent with QCD.
In terms of effective theories this means that they should match to QCD at a scale
close to the chiral scale $\Lambda_\chi \sim 4\pi f_\pi \sim 1$ GeV.
It has been shown that this matching can be effectuated in the framework
of hidden local symmetry (HLS) and leads to what is called `vector manifestation' (VM)ве
~\cite{Harada:2003jx} which provides a theoretical support  for a low-energy effective
field theory for hadrons and which gives, in the chiral limit, an elegant and unambiguous
prediction of the behavior of light-quark hadrons at high temperature and/or at high
density. Following the indications of the  HLS theory,  we have described a Skyrme model
in which the dilaton field $\chi$, whose role in dense matter was first pointed out by
Brown and Rho,~\cite{Brown:1991kk} and the vector meson fields $\rho$ and $\omega$ were
incorporated into the Skyrme Lagrangian to construct dense skyrmion matter.

We have presented an approach to hadronic physics based on Skyrme's philosophy, namely
that baryons are solitons of a theory described in terms of meson fields,
which can be justified from QCD in the large $N_c$ expansion.
We have adopted  the basic principles of effective field theory.
Given a certain energy domain we describe the dynamics by a Lagrangian defined
in terms of the mesonic degrees of freedom active in that domain, we thereafter implement
the symmetries of QCD and VM, and describe the baryonic sectors
as topological winding number sectors and solve in these sectors the equations derived from the Lagrangian
with the appropriate  boundary conditions for the sector.
In this way one can get all of Nuclear Physics out of a single Lagrangian.
We have studied  the B=1 sector to obtain the properties of the single skyrmion,
the B=2 sector to understand the interaction between skyrmions, and our main effort has been
the study skyrmion matter, as a model for hadronic matter, investigating its behavior
at finite density and temperature and the description of meson properties in that dense medium.

Skyrme models have been proven successful in describing nuclei, the nucleon-nucleon interaction and
pion-nucleon interactions. It turns out that Skyrme models also represent a nice tool for understanding
low density cold hadronic matter and the behavior of  the mesons in particular the pion inside matter.
We have shown in here that when hadronic matter is compressed and/or heated  Skyrme models
provide useful information on the chiral phase transitions. Skyrmion matter is realized as a crystal and we have seen
that at low densities it is an FCC crystal made of skyrmions. The phase transition occurs when the FCC crystal transforms
into a half skyrmion CC one. In our study we have
discovered  the crucial role of the scale dilaton in describing the expected phase transition towards a chiral symmetry restored phase.
We have also noticed the peculiar behavior of the $\omega$ associated to its direct coupling to the baryon number
current and we have resolved the problem by naturally scaling the coupling constant
using the scale dilaton.

Another aspect of our review has been the study of the properties of elementary mesons in the medium,
in particular those involved in the model, the pion and the dilaton. Moreover we have described how their properties
 change when we move from one phase to another.
 
A description of the chiral restoration phase
transition in the temperature-density plane has been presented, whose main ingredient
is that the dominant scenario is the absorption of heat by the fluctuating pions in the
background of crystal skyrmion matter. This description
leads to a phase transition whose dynamical structure  is parameter independent
and whose shape resembles much the conventional confinement/deconfiment phase transition.
We obtain, for parameter values close the conventional ones,
the expected critical temperatures and densities.

For clarity, the presentation has been linear, in the sense, that given the Lagrangian we  have described its phenomenology,
and have made no effort to interpret the mechanisms involved and the results obtained from QCD. In this way we have taken a `bottom up' approach:
the effective theory represents confined QCD and it should explain the hadronic phenomenology in its domain of validity.

The main result of our calculation is the realization that phase transition scenario is not as simple as initially thought
but contains many features which make interesting and phenomenologically appealing. It is now time to try to collect
ideas based on fundamental developments and see how our effective theory and the principles
that guide it realize these ideas.  In this line of thought, it is exciting to have unveiled  scenarios near the phase transition
 of unexpected interesting phenomenology
in line with recent proposals.~\cite{McLerran:2008ux,McLerran:2007qj}

\section*{Acknowledgements}
We would like to thank our long time collaborators  Dong-Pil Min and Hee-Jung Lee
whose work is reflected in these pages and who have contributed greatly to the effort.
We owe inspiration and gratitude to Mannque Rho, who during many years has been a
motivating force behind our research.  Skyrmion physics had a boom in the late 80's and
thereafter only a few groups have maintained this activity obtaining very beautiful results,
which however, have hardly influenced the community.
We hope that this book contributes to make skyrmion physics more widely appreciated.
Byung-Yoon Park thanks the members of  Departamento de F\'{\i}sica Te\'orica of the University
of Valencia for their hospitality. Byung-Yoon Park and Vicente Vento were supported by
 grant FPA2007-65748-C02-01 from Ministerio de Ciencia e Innovaci\' on.

\bibliographystyle{ws-rv-van}
\bibliography{byp}

\printindex                         % to print subject index

\end{document}